\documentclass[sn-nature]{sn-jnl}

\usepackage{placeins}

\newcommand{\pasp}{Publ. Astron. Soc. Pac.}
\newcommand{\apjl}{Astrophys. J. Lett.}
\newcommand{\apj}{Astrophys. J.}
\newcommand{\aj}{Astron. J.}
\newcommand{\apjs}{Astrophys. J. Suppl. Ser.}
\newcommand{\mnras}{Mon. Not. R. Astron. Soc.}
\newcommand{\nat}{Nature}
\newcommand{\aap}{Astron. Astrophys.}

\newcommand{\arcsec}{$"$}
\newcommand{\icarus}{Icarus}
\newcommand{\aaps}{Astron. Astrophys. Suppl.}
\newcommand{\jatis}{J. Astron. Telesc. Instrum. Syst.}
\newcommand{\jve}{J. Vib. Eng.}
\newcommand{\joss}{J. Open Source Softw.}
\newcommand{\rnaas}{Res. Notes AAS}

\usepackage[title]{appendix}%
\usepackage{textcomp}%
\usepackage{float}%
\usepackage{fancyhdr}

\usepackage{graphicx}%
\usepackage{multirow}%
\usepackage{amsmath,amssymb,amsfonts}%
\usepackage{amsthm}%
\usepackage{mathrsfs}%
\usepackage[title]{appendix}%
\usepackage{xcolor}%
\usepackage{textcomp}%
\usepackage{manyfoot}%
\usepackage{booktabs}%
\usepackage{algorithm}%
\usepackage{algorithmicx}%
\usepackage{algpseudocode}%
\usepackage{listings}%

\newcommand{\mjup}{\ensuremath{M_{\rm Jup}}}

\newcommand{\rpl}{\ensuremath{R_{p}}}

\newcommand{\rstar}{\ensuremath{R_\star}}
\newcommand{\mstar}{\ensuremath{M_\star}}

\newcommand{\teffstar}{\ensuremath{T_{\rm eff\star}}}
\newcommand{\rhostar}{\ensuremath{\rho_\star}}
\newcommand{\loggstar}{\ensuremath{\log{g_{\star}}}}

\newcommand{\arstar}{\ensuremath{a/\rstar}}

\newcommand{\ecosw}{\ensuremath{\sqrt{e}\cos\omega}}
\newcommand{\esinw}{\ensuremath{\sqrt{e}\sin\omega}}

\newcommand{\gcmc}{\ensuremath{\rm g\,cm^{-3}}}

\newcommand{\starKOIID}{\ensuremath{134}}
\newcommand{\starKICID}{\ensuremath{9032900}}
\newcommand{\starTICID}{\ensuremath{271772050}}
\newcommand{\starGaiaID}{\ensuremath{2079971473693847168}}
\newcommand{\starRA}{19:42:20.78}
\newcommand{\starDec}{45:21:53.77}
\newcommand{\starRefEpoch}{\ensuremath{2016}}
\newcommand{\starPMRA}{\ensuremath{-2.264\pm0.012}}
\newcommand{\starPMDec}{\ensuremath{-6.266\pm0.013}}

\newcommand{\starParallax}{\ensuremath{0.870\pm0.010}}
\newcommand{\starTMag}{\ensuremath{13.2243\pm0.0088}}
\newcommand{\starKepMag}{\ensuremath{13.675}}
\newcommand{\starGaiaMag}{\ensuremath{13.6195\pm0.00032}}
\newcommand{\starGaiaBPMag}{\ensuremath{13.9226\pm0.0014}}
\newcommand{\starGaiaRPMag}{\ensuremath{13.1573\pm0.00081}}
\newcommand{\starJMag}{\ensuremath{12.628\pm0.023}}
\newcommand{\starHMag}{\ensuremath{12.413\pm0.021}}
\newcommand{\starKMag}{\ensuremath{12.322\pm0.018}}
\newcommand{\starVMag}{\ensuremath{13.773\pm0.137}}
\newcommand{\starBMag}{\ensuremath{14.233\pm0.07}}

\newcommand{\ticMass}{\ensuremath{1.17}}
\newcommand{\ticRadius}{\ensuremath{1.81}}
\newcommand{\ticTeff}{\ensuremath{6153\pm26}}

\newcommand{\starDistance}{\ensuremath{1.118_{-0.014}^{+0.022}}}
\newcommand{\starMass}{\ensuremath{1.41\pm.14}}
\newcommand{\starRadius}{\ensuremath{1.73\pm0.25}}

\newcommand{\starLuminosity}{\ensuremath{4.2_{-1.2}^{+1.1}}}
\newcommand{\starTeff}{\ensuremath{6160\pm130}}

\newcommand{\bUpperMSinI}{\ensuremath{9}}

\newcommand{\starRho}{\ensuremath{1.32\pm0.66}}
\newcommand{\starLogg}{\ensuremath{4.11\pm0.13}}
\newcommand{\starKepleruOne}{\ensuremath{0.353_{-0.036}^{+0.039}}}
\newcommand{\starKepleruTwo}{\ensuremath{0.287\pm.075}}

\newcommand{\bPeriod}{\ensuremath{67.5258}}
\newcommand{\bROR}{\ensuremath{0.06618_{-0.00043}^{+0.00046}}}
\newcommand{\bImpactParameter}{\ensuremath{0.386_{-0.135}^{+0.076}}}
\newcommand{\bSemimajorAxis}{\ensuremath{0.3653_{-0.0096}^{+0.0088}}}
\newcommand{\bAOR}{\ensuremath{46.2_{-4.5}^{+5.5}}}
\newcommand{\bInclination}{\ensuremath{89.52_{-0.13}^{+0.19}}}
\newcommand{\bRadius}{\ensuremath{1.074\pm0.017}}

\newcommand{\bTeq}{\ensuremath{659_{-46}^{+35}}}

\newcommand{\bIrr}{\ensuremath{45_{-10}^{+11}}}
\newcommand{\bDuration}{\ensuremath{11.3_{-1.2}^{+0.96}}}
\newcommand{\bIngressDuration}{\ensuremath{67_{-18}^{+15}}}

\unnumbered

\begin{document}

\title[Article Title]{A high mutual inclination system around KOI-134 revealed by transit timing variations}

\author*[1]{\fnm{Emma} \sur{Nabbie}}\email{Emma.Nabbie@usq.edu.au}

\author[1]{\fnm{Chelsea X.} \sur{Huang}}\nomail

\affil[1]{\orgdiv{Centre for Astrophysics}, \orgname{University of Southern Queensland}, \orgaddress{\street{West St}, \city{Toowoomba}, \postcode{4350}, \state{Queensland}, \country{Australia}}}

\author[2,3]{\fnm{Judith} \sur{Korth}}\nomail

\affil[2]{\orgdiv{Lund Observatory, Division of Astrophysics, Department of Physics}, \orgname{Lund University}, \orgaddress{\street{Box 118}, \city{Lund}, \postcode{22100}, \country{Sweden}}}
\affil[3]{\orgdiv{Observatoire Astronomique de l’Universit\'{e} de Gen\`{e}ve}, \orgaddress{\street{Chemin Pegasi 51}, \city{Versoix}, \postcode{1290}, \country{Switzerland}}}

\author[4,5]{\fnm{Hannu} \sur{Parviainen}}
\affil[4]{\orgdiv{Departamento de Astrof{\'i}sica}, \orgname{Universidad de La Laguna (ULL)}, \orgaddress{\street{38206 La Laguna}, \city{Tenerife}, \country{Spain}}}
\affil[5]{\orgname{Instituto de Astrof\'{i}sica de Canarias (IAC)}, \orgaddress{\street{38200 La Laguna}, \city{Tenerife}, \country{Spain}}}

\author[6,7]{\fnm{Su} \sur{Wang}}
\affil[6]{\orgdiv{CAS Key Laboratory of Planetary Sciences}, \orgname{Purple Mountain Observatory, Chinese Academy of Sciences}, \city{Nanjing}, \postcode{210023}, \country{People’s Republic of China}}
\affil[7]{\orgname{CAS Center for Excellence in Comparative Planetology}, \city{Hefei}, \postcode{230026}, \country{People’s Republic of China}}

\author[1]{\fnm{Alexander} \sur{Venner}}\nomail

\author[1]{\fnm{Robert} \sur{Wittenmyer}}\nomail

\author[8]{\fnm{Allyson} \sur{Bieryla}}
\affil[8]{\orgdiv{Center for Astrophysics}, \orgname{Harvard and Smithsonian}, \orgaddress{\street{60 Garden Street}, \city{Cambridge}, \postcode{02138}, \state{Massachussetts}, \country{USA}}}

\author[8]{\fnm{David W.} \sur{Latham}}

\author[9]{\fnm{Gongjie} \sur{Li}}
\affil[9]{\orgdiv{Center for Relativistic Astrophysics, School of Physics}, \orgname{Georgia Institute of Technology}, \orgaddress{\city{Atlanta}, \postcode{30332}, \state{Georgia}, \country{USA}}}

\author[10,11]{\fnm{Douglas N. C.} \sur{Lin}}
\affil[10]{\orgdiv{Department of Astronomy and Astrophysics}, \orgname{University of California, Santa Cruz}, \orgaddress{\city{Santa Cruz}, \postcode{95064}, \state{California}, \country{USA}}}
\affil[11]{\orgdiv{Institute for Advanced Studies}, \orgname{Tsinghua University}, \orgaddress{\city{Beijing}, \postcode{100086}, \country{People’s Republic of China}}}

\author[1]{\fnm{George} \sur{Zhou}}

\abstract{Few planetary systems have measured mutual inclinations, and even less are found to be non-coplanar. Observing the gravitational interactions between exoplanets is an effective tool to detect non-transiting companions to transiting planets. Evidence of these interactions can manifest in the light curve through transit timing variations (TTVs) and transit duration variations (TDVs). Through analysis of \textit{Kepler} photometry and joint TTV-TDV modeling, we confirm the detection of KOI-134 b, a transiting planet with mass and size similar to Jupiter on a period of $\sim$67 days, and find that it exhibits high TTVs (~20-hr amplitude) and significant TDVs. We explain these signals with the presence of an innermost non-transiting planet in 2:1 resonance with KOI-134\,b. KOI-134\,c has a mass $M = 0.220^{+0.010}_{-0.011} M_\text{Jup}$ and a moderately-high mutual inclination with KOI-134\,b of {$i_\text{mut} = 15.4_{-2.5}^{+2.8}{^\circ}$}. Moreover, the inclination variations of KOI-134\,b are so large that the planet is predicted to stop transiting in about 100 years. This system architecture cannot be easily explained by any one formation mechanism, with other dynamical effects needed to excite the planets' mutual inclination while still preserving their resonance.}


\maketitle

\section*{Main}

The mutual inclination of a planetary system -- the angle between the orbital planes of neighboring planets -- encodes information about its formation history. However, measuring mutual inclinations requires precise knowledge about the orbital elements of exoplanets, information which is inaccessible in many cases. 

Of the few systems with direct estimates of their mutual inclinations, most are coplanar ($\lesssim 5^{\circ}$) (e.g. GJ 876 \cite{Rivera:2010}; Kepler-30 \cite{Sanchis-Ojeda:2012}; KOI-872 \cite{Nesvorny:2012}; Kepler-56 \cite{Huber:2013}; Kepler-119 \cite{Almenara:2015}; K2-146 \cite{Hamann:2019}; V1298 Tau\,b \cite{Johnson:2022} and c \cite{Feinstein:2021}). These systems are thought to be produced by dynamically-quiet processes like disk migration, which should produce low mutual inclinations \cite{GoldreichTremaine:1980}. The results of these individual studies align with the statistical estimations of the mutual inclination distribution of \textit{Kepler} systems with multiple transiting planets (e.g. \cite{Lissauer:2011}). These statistical analyses found that \textit{Kepler} systems on average had mutual inclinations between 1-2$^{\circ}$, after modeling the distribution of transit duration ratios between planet pairs \cite{Steffen:2010, Fang:2013, Fabrycky:2014}. Therefore, the standard \textit{Kepler} multi-planet system tends to have an architecture similar to our own solar system planets, whose inclinations do not exceed {$\sim3^{\circ}$} on average (excluding Mercury).

Systems with high mutual inclinations ($\gtrsim 10^{\circ}$) are nevertheless expected to be observed, as processes like planet-planet scattering are theorized to excite mutual inclinations (e.g. \cite{Marzari:2002, Chatterjee:2008, JuricTremaine:2008}). A handful of planets in radial velocity systems like HD 147018 and HD 168443 are also estimated to have high mutual inclinations through their sky-projected apsidal separations \cite{DawsonChiang2014}. Astrometric measurements of Upsilon Andromedae \cite{McArthur:2010} revealed a {$\sim$30$^\circ$} mutual inclination. To date, only two systems with transiting planets are found to have long period, high mutually-inclined, non-transiting companions: Kepler-419 ($9^{+8}_{-6}$$^\circ$ \cite{Dawson:2014}) and Kepler-108 ($24.2^{+10.8}_{-7.8}$$^\circ$ \cite{MillsFabrycky:2017}). This paucity is in part due to the fact that higher mutual inclinations decrease the probability that all planets in a system are transiting, adding the additional challenge of obtaining dynamical information for bodies we may not be able to observe. 

However, we can overcome this in special cases, where dynamical perturbations of a companion cause transit timing variations (TTVs) in a transiting planet. This allows us to precisely characterize the three-dimensional architecture of a multi-planet system through the dynamical behavior of its planets. 
Such is the case with KOI-134\,b, a transiting warm Jupiter that was originally discarded as a false positive by the \textit{Kepler} pipeline due to its 20-hour TTVs (Fig. 1, Fig. 2a).

\subsection{Results}

We report the discovery of a transiting planet with a non-transiting planet companion around KOI-134, an F-type star ($V$ = 13.8), via TTVs. KOI-134 has an effective temperature of 6160\,K, with a mass of 1.41\,{$M_\odot$} and a radius of 1.73\,{$R_\odot$}. Through analysis of archival radial velocity and direct imaging data, we find that KOI-134 does not have any stellar-mass companions that would cause the transit signals observed in the 17 quarters of Kepler data (Extended Data Fig. \ref{fig:directimage}; see Methods for further details).

The TTVs of KOI-134\,b are caused by the gravitational influence of another body in the system, thus showing that KOI-134\,b is indeed a planet, and it is accompanied by a non-transiting planet that we will refer to as KOI-134\,c. Along with TTVs, KOI-134\,b also exhibits transit duration variations (TDVs; Fig. 2b), indicative of a changing impact parameter. These TTVs and TDVs allow us to measure the masses of both planets and gain a comprehensive view of the KOI-134 system architecture.

The best-fit solution from our TTV-TDV dynamical modeling is shown in Table \ref{tab:nbody}. Results of our per-epoch modeling of the transit times and transit durations using the Kepler light curves (see Methods for details) are summarized in Extended Data Table \ref{tab:fit}. Individual transits overlaid with our best-fit transit model and associated residuals are shown in Supplementary Fig. 1. We then compute the TTVs relative to a linear ephemeris ($P$ = 67.53 days, $t_0$ = 2454917.9306). The transit durations of KOI-134\,b slightly increased during the first two thirds of the Kepler mission, and then decreased towards the end of the mission. The maximum difference between the durations of two transits are approximately 20 minutes. Overall, the measured transit durations deviates from a constant transit duration model by {$\sim$}6{$\sigma$}.

Through joint TTV-TDV modeling with N-body simulations using {\tt REBOUND} \cite{rebound} (see Methods), we find that the signals can be explained by a non-transiting planet at the inner 2:1 resonance with respect to KOI-134\,b. Our best fit results show that KOI-134\,b has a mass of $1.09^{+0.12}_{-0.08}$ {$M_\text{Jup}$} and an orbital period of $67.1277^{+0.0045}_{-0.0057}$ days. Moreover, we find that KOI-134\,c is a sub-Saturn, with a mass of $0.220^{+0.010}_{-0.011}$ {$M_\text{Jup}$} and an orbital period of $33.950^{+0.013}_{-0.020}$ days. We find that both planets are modestly eccentric, with KOI-134\,b having a eccentricity of $0.16^{+0.02}_{-0.03}$ and KOI-134\,c having an eccentricity of $0.24^{+0.12}_{-0.03}$. We also performed independent photodynamic modeling following procedures described in \cite{Korth:2023, Korth:2024} and arrived at similar best fit system parameters. High-precision N-body parameters for our adopted solution are summarized in Extended Data Table \ref{tab:nbodyhp}.

We find that a moderately high mutual inclination ({$i_\text{mut} = 15.4_{-2.5}^{+2.8}{^\circ}$}) between KOI-134\,b and KOI-134\,c is required to explain the observed TTV and TDV signals. This is a clear outlier from the average mutual inclination range of 1.0{$^\circ$}-2.2{$^\circ$} for \textit{Kepler} multi-planet systems \cite{Fabrycky:2014}. A schematic drawing of the KOI-134 system architecture in comparison to the coplanar, ``solar system like" configuration of planets is shown in Fig. \ref{fig:schematic}. KOI-134 is the only system with planets that are significantly mutually inclined and near a first-order mean motion resonance (MMR) to date.

We further use {\tt REBOUND} to examine the long term stability and evolution of the system. We found that the system is stable through 10 Myrs of evolution. The mutual inclination between the two planets drives orbital procession at time scales of $\approx$ 800 years (Fig. 4h). The mutual inclinations between the planets varies between 1.2{$^\circ$}-39.5{$^\circ$}. However, while KOI-134\,b is still transiting now, it only spends about 20\% of its lifetime transiting its host star from our line of sight. Fig. 4c shows that the inclination of KOI-134\,b is predicted to lower such that the planet will no longer transit its host star. The planet's transit signal is expected to disappear completely around the year 2059. Our orbit evolution predicts that KOI-134\,c will never transit despite its changing orbit inclination angle.

KOI-134 is also observed by the \textit{Transiting Exoplanet Survey Satellite} (\textit{TESS}, \cite{ricker}) in the Full Frame Images for six total sectors across 5 years. We identified one candidate transit event in the \textit{TESS} sector 54 data that matches the predicted transit times from the posteriors of \textit{Kepler} TTV modeling (see Methods). However, due to the relatively low signal to noise of the event, we cannot use this event to refine our dynamic solutions.   

\subsection{Discussion}

KOI-134\,b has one of the highest-amplitude TTVs observed (Extended Data Fig. \ref{fig:ttvdist}), and the KOI-134 system lies well within the resonant domain. This is one of the few systems within the 2:1 resonant domain, such as TOI-216 \cite{Dawson:2019,Dawson:2021,Nesvorny:2022}, whereas other systems like KOI-142 \cite{Nesvorny:2013} can still lie outside of the resonant domain despite their large TTVs. With a period ratio of 1.9772, the system has a normalized distance to the 2:1 MMR of {$\Delta$= -0.011}, which indicates that the system lies narrow of the 2:1 resonance \cite{Lithwick2012}. We show the position of the KOI-134 system within the resonant domain parameterized by {$X$} and {$\Gamma$} \cite{Deck:2013} and compare its position to other TTV systems near the 2:1 MMR (Extended Data Fig. \ref{fig:resonance}).  

We can determine that KOI-134\,b and KOI-134\,c are librating in resonance by examining the evolution of their eccentricity resonant angles, {$\Theta_{e,b}$} and {$\Theta_{e,c}$}. These angles are defined as follows:
\begin{equation}
\Theta_{e,b} = 2\lambda_{b} - \lambda_{c} - \varpi_{b},
\end{equation}
\begin{equation}
\Theta_{e,c} = 2\lambda_{b} - \lambda_{c} - \varpi_{c},
\end{equation}
where $\lambda_i = M_i + \varpi_i$ and $\varpi_i = \omega_i + \Omega_i$ \cite{Murray:1999}.

A segment of the orbital evolution across 10000 years is depicted in Fig. 4. The resonant angle of the inner planet ({$\Theta_{e,c}$}) shows libration about 0 with a libration period of $\sim$4.5 years. In contrast, the resonant angle of the outer, transiting planet ({$\Theta_{e,b}$}) shows no libration. Extended Data Fig. \ref{fig:trajectory} also shows KOI-134\,b and KOI-134\,c's trajectories librate around a fixed point. Additionally, we found that the inclination resonant angles corresponding to the 4:2 inclination resonance do not show any libration for both planets. 

KOI-134 occupies a unique niche in terms of system architecture. While it joins many other near-MMR, giant-hosting multi-planetary systems (Extended Fig. \ref{fig:population}), it is the only one to host a planet pair with a moderately high mutual inclination (1.2{$^\circ$}-39.5{$^\circ$} through a 10 Myr evolution). For example, the pair of planets in the TOI-216 system are similarly in a 2:1 resonance, yet only have a mutual inclination between 1.2-3.9$^{\circ}$. 

One possible way to excite the mutual inclination of the KOI-134 system is through resonant inclination excitation \cite{ThommesLissauer:2003}, where the planets first pass through the 4:2 inclination resonance during disk migration. We perform preliminary simulations using the methodology from \cite{Cresswell:2008} and assume various migration timescales to investigate different recipes for disk migration. In these simulations, and under our eccentricity and inclination damping assumptions, we find that in order to excite the mutual inclination to 10 degrees or above, the eccentricities of both planets also need to be excited to much larger values ($\approx$ 0.5) than our observational constraints. This joint eccentricity and inclination excitation matches the results shown in Fig. 2 of \cite{ThommesLissauer:2003}. 

KOI-134 therefore presents an interesting test case of planet formation theories. The observed TTVs and TDVs indicate that these planets were captured in a resonance while the disk was still present, with the high mutual inclination suggesting perturbation of the planets' orbits at later stages of formation. Further investigation is needed to disentangle the various mechanisms needed to produce the observed mutual inclination yet preserve the strong resonance that the planets currently maintain.

\begin{figure*}[htp]
\centering
\includegraphics[width=1\textwidth]{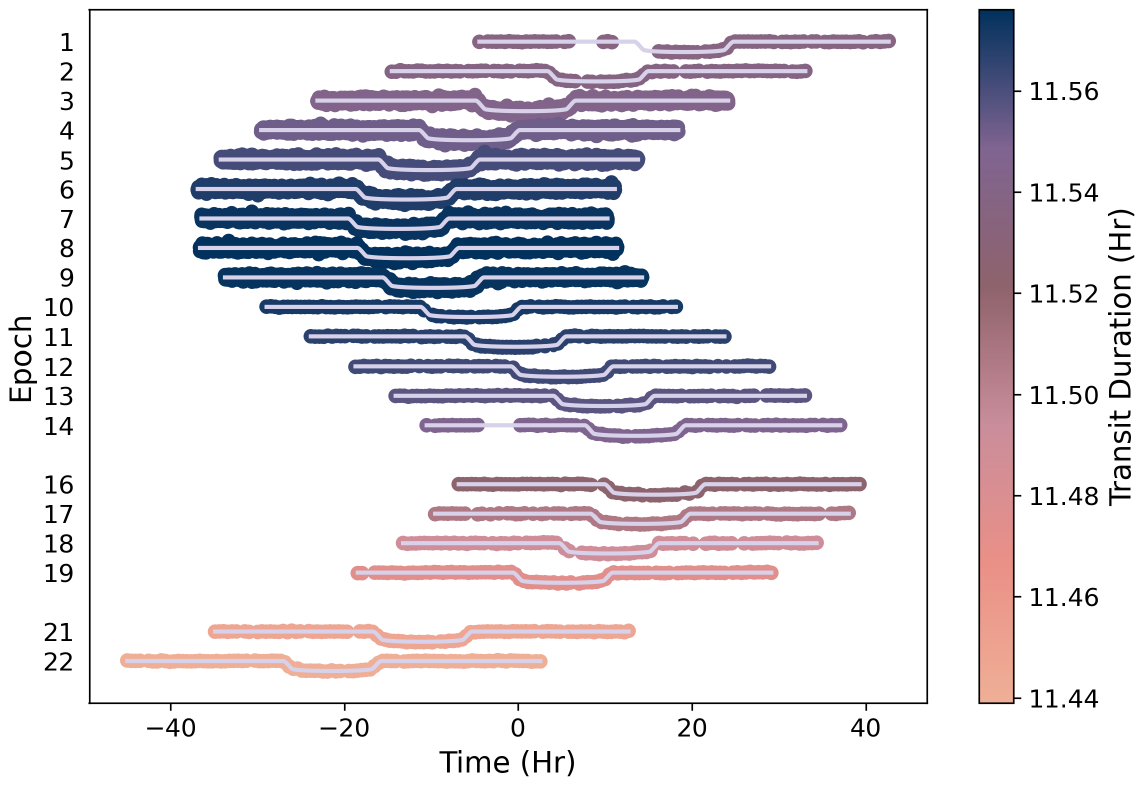}\hfill

\caption{\textbf{Per-epoch Kepler light curves of KOI-134\,b.} {Least-squares detrended \textit{Kepler} light curves, with a {\tt batman} model overplotted in grey. The transit epochs are separated by a trivial offset value in flux simply for viewing purposes. The horizontal axis represents the time offset between the observed and the calculated (O-C) transit time, where the latter time was derived from assuming a linear orbital period and extrapolating from the first observed transit epoch.}}
\label{fig:riverplot}

\end{figure*}

\begin{figure*}[htp]
\centering
\includegraphics[width=1\textwidth]{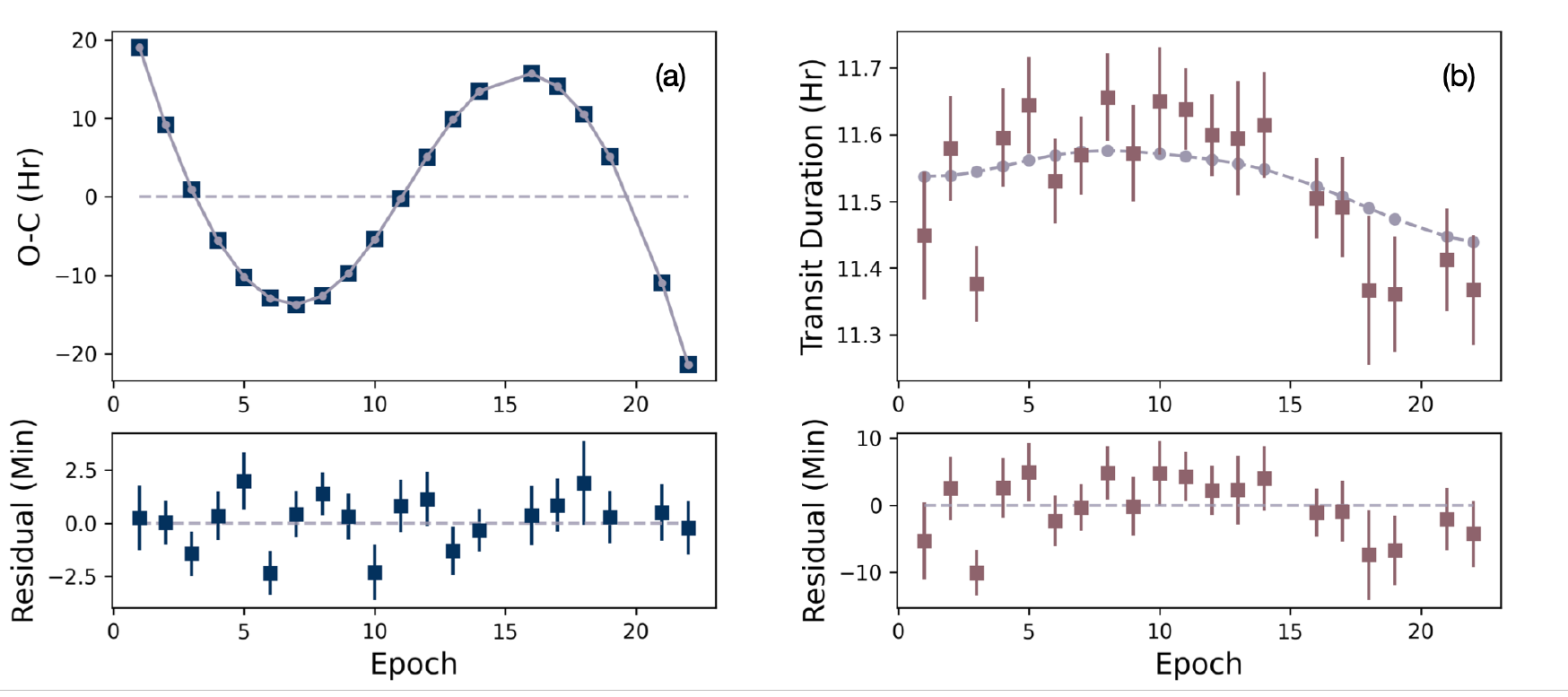}\hfill

\caption{\textbf{Transit timing and transit duration variations of KOI-134\,b.} \textit{Panel (a):} Time as a function of the 20 observed minus calculated (O-C) transit times based on a linear ephemeris. Modeled TTVs from $N$-body simulations are shown in grey, while observed TTVs are shown in blue. The residuals are shown in the panel directly below. \textit{Panel (b):} Transit duration as a function of epoch. The 20 observed transit durations are shown in brown, while modeled transit durations from {\tt REBOUND} are denoted in grey. The residuals are located in the panel below. All panels include 1$\sigma$ errorbars with points derived from the median posterior spread.}
\label{fig:ocplot}

\end{figure*}

\begin{figure*}[htp]
\centering
\includegraphics[width=1\textwidth]{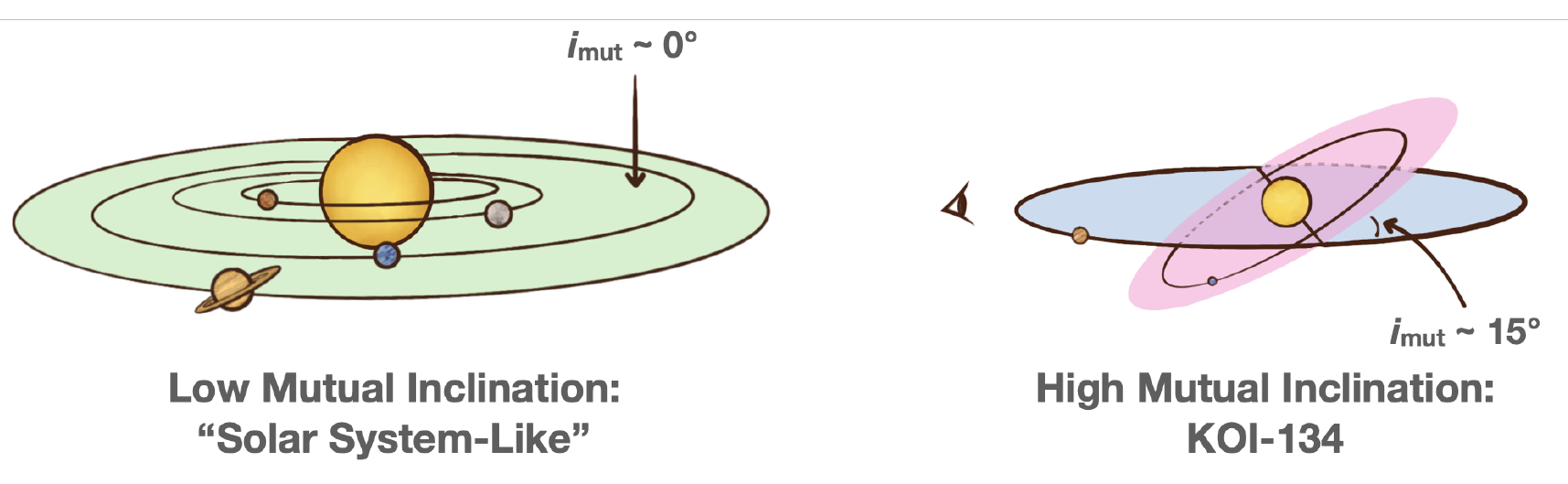}\hfill

\caption{\textbf{Visualization of high versus low mutual inclination systems}. Systems with low mutual inclinations (left) are more reminiscent of the solar system, whereas the planets around KOI-134 (right) deviate strongly from coplanarity. The angle between the pink and blue orbital planes, $i_\text{mut}$, denotes the mutual inclination of the KOI-134 system.}
\label{fig:schematic}

\end{figure*}

\begin{figure*}[htp]
\centering
\includegraphics[width=0.9\textwidth]{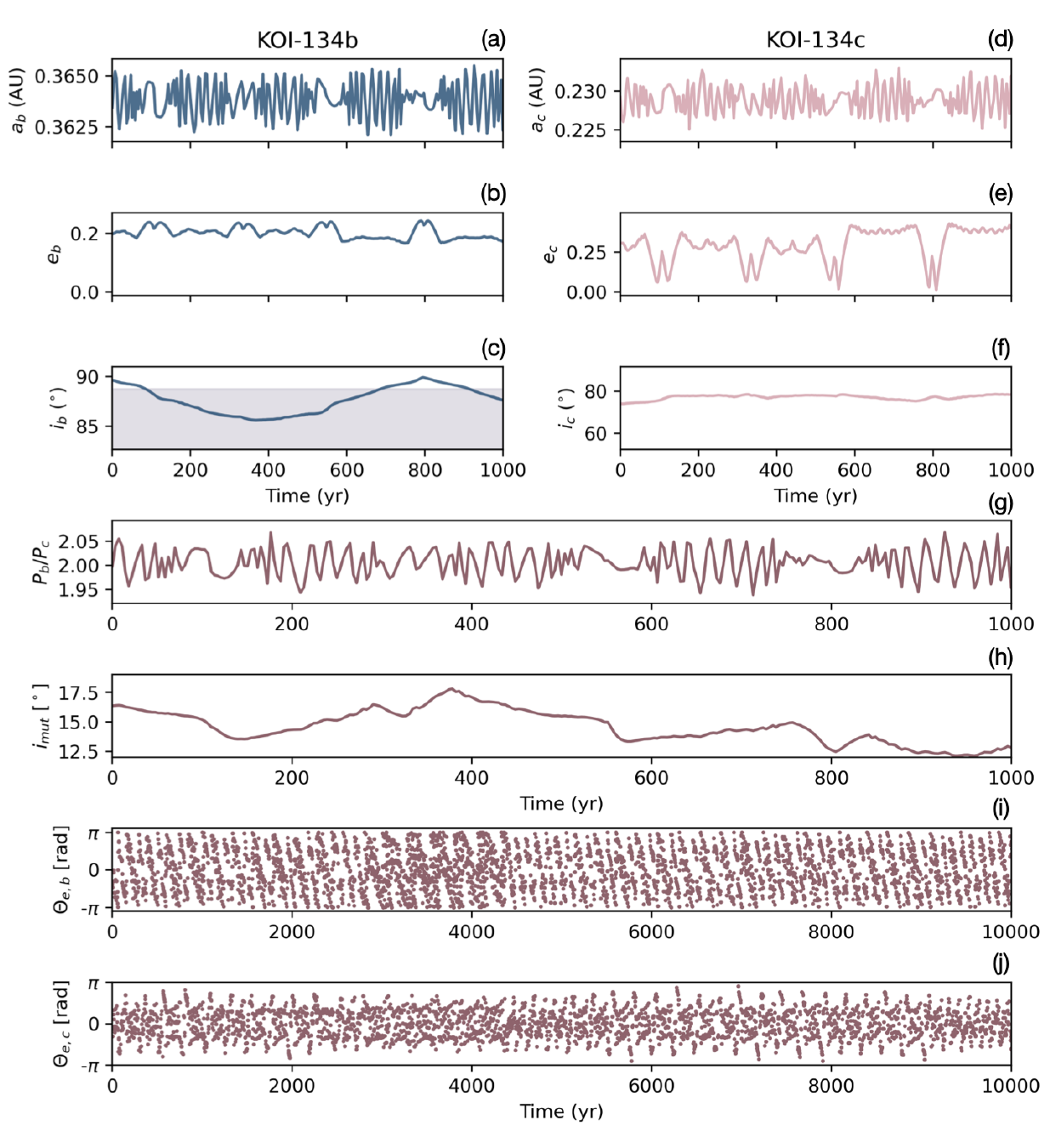}\hfill

\caption{\textbf{Evolution of the KOI-134 system architecture}. Orbital elements over time for KOI-134b (blue) and KOI-134c (pink) integrated using the {\tt REBOUND} ias15 integrator with general relativity effects included. The top five rows show the first 1000 years of our integration that spans 10 Myr in total. The bottom two rows extend to 10000 years to show the long-term behavior of the resonant angles. {\textit{Panels (a)-(f):}} Evolution of semimajor axis (top), eccentricity (middle), and inclination (bottom) of both planets. The shaded region in panel (c) denotes inclinations which would prevent KOI-134\,b from transiting. \textit{Panels (g)-(j):} Period ratio {$P_b$/{$P_c$}}, mutual inclination {$i_\text{mut}$}, and resonant angles {${\Theta}_{e,b}$}, {${\Theta}_{e,c}$} as a function of time.}
\label{fig:evolution}

\end{figure*}

\begin{table}[h]
\caption{
  Best-Fit System Parameters from $N$-body Simulations
  \label{tab:nbody}
}
\begin{tabular}{lcc}
\toprule
Parameter & Non-Photodynamic\footnotemark[1] & Photodynamic \\
\midrule
\textbf{KOI-134\,b} & & \\
Mass [{\mjup}] & $1.09^{+0.12}_{-0.08}$ & $1.00^{+0.25}_{-0.27}$ \\
Period [days] & $67.1277^{+0.0045}_{-0.0057}$ & $67.701^{+0.016}_{-0.017}$ \\
Eccentricity & $0.16^{+0.02}_{-0.03}$ & $0.05^{+0.022}_{-0.020}$ \\
Argument of of periastron, {$\omega$} [$^\circ$] & $153.2^{+10}_{-5}$ & $227^{+12}_{-13}$ \\
Longitude of ascending node, {$\Omega$} [$^\circ$] & $33.9^{+6.8}_{-3.3}$ & $179.9948^{+0.0056}_{-0.0054}$\\
Inclination [$^\circ$] & $89.58^{+0.13}_{-0.093}$ &  90.501 {$\pm$} 0.043 \\
Mean anomaly, {$M$} [$^\circ$] &  $2.2^{+3.5}_{-1.1}$ & $201^{+14}_{-13}$ \\
Semimajor axis, {$a$} [AU] & $0.363^{+0.011}_{-0.013}$ & $0.364637^{+5.1e-5}_{-6.4e-5}$  \\
[1.5ex]
\textbf{KOI-134\,c} & & \\
Mass [{\mjup}] & $0.22^{+0.017}_{-0.021}$ & $0.257^{+0.046}_{-0.036}$ \\
Period [days] & $33.950^{+0.013}_{-0.020}$ & 32.89 {$\pm$} 0.11 \\
Eccentricity & $0.24^{+0.12}_{-0.03}$ & $0.179^{+0.025}_{-0.022}$ \\
Argument of periastron, {$\omega$} [$^\circ$] & $-13^{+16}_{-5}$ & $-116.6^{+7.9}_{-7.4}$ \\
Longitude of ascending node, {$\Omega$} [$^\circ$] & $224^{+12}_{-2}$ & $176.7^{+2.5}_{-2.6}$ \\
Inclination [$^\circ$] & $75.0^{+2.9}_{-2.7}$ & $75.9^{+4.0}_{-5.1}$ \\
Mean anomaly, {$M$} [$^\circ$] & $-2.85^{+5.3}_{-3.9}$ & $-0.74^{+0.30}_{-0.32}$ \\
Semimajor axis, {$a$} [AU] & $0.2305^{+0.0069}_{-0.0081}$ & $0.22534^{+0.00052}_{-0.00050}$ \\
\botrule
\end{tabular}
\footnotetext[1]{We adopt these parameters as our best-fit solution.}
\end{table}

\section*{Methods}\label{sec:methods}
\subsection{Data}
\subsubsection{\textit{Kepler} data and dispositional history}\label{subsec2}
\textit{Kepler} time-series photometry of KOI-134 (KIC~9032900) was taken from May 2009 to April 2013. Long-cadence (LC) photometric data is available for Quarters 1 to 17, while short-cadence (SC) observations are available from Quarter 3 to Quarter 7. The \textit{Kepler} Science Operations Pipeline detected a 67.686-day transit signal with a signal-to-noise ratio of 11.7 and Multi Event Statistics  of 97.1 \citep{jenkins2010}. Upon visual inspection, the transits detected by the \textit{Kepler} pipeline do not align with the actual transit events due to the strength of the candidate's TTVs. These strong offsets in expected versus observed period caused the signals of one planet to be misidentified as four separate planet candidates, though only the 67-day signal was eventually reported. The pipeline found neither significant out-of-transit centroid offsets nor secondary eclipses, indicating that there are no stellar companions in the KOI-134 system.

The transiting planet, KOI-134\,b, was first alerted as a KOI in Q1 (June 2009). However, the system failed the \textit{Kepler} Robovetter's \citep{Coughlin:2016} Model-Shift Uniqueness Test due to the transiting planet's strong TTVs. As a result, KOI-134\,b was dispositioned as a false positive \citep{Borucki:2011}, and retained this label up to DR25 \citep{Thompson:2018}. Through the efforts of the \textit{Kepler} False Positive Working Group (FPWG; \cite{Bryson:2017}), KOI-134\,b was reversed to candidate status in the DR25 Supplemental KOI Table. The procedures for rescuing misidentified planet candidates are outlined in \cite{Vanderburg:2020}.

\subsubsection{Reconnaissance spectra}\label{subsec3}
Archival data from the Tillinghast Reflector Echelle Spectrograph (TRES; \cite{TRES}) provided nine high-resolution spectral observations of KOI-134 (Supplementary Table 1). These observations span a wavelength range of 3850 to 9100 {\AA} and were taken from November 2012 to July 2013. Each exposure lasted between 2700 and 3000 seconds, yielding SNRs that range between 24.4 to 30.2. The TRES instrument is located at the Fred L. Whipple Observatory in Mount Hopkins, Arizona. Due to the large jitter in the measurements, we cannot constrain the mass of KOI-134\,b. We report two radial velocity solutions, derived from both Cross-Correlation Function (CCF) as well as Least-Squares Deconvolution (LSD; \citep{ZhouLSD}) techniques. We also use the LSD technique to derive a stellar $v\text{sin}(i)$ of 10.59 $\pm$ 0.38 km s$^{-1}$, which contributes to the large jitter.

KOI-134 was also observed on UT 2012-01-02 by the High-Resolution Echelle Spectrometer (HIRES; \cite{HIRES}) instrument at Keck Observatory's 10-meter telescope as part of the California Kepler Survey. The observations had a signal-to-noise ratio of 44 and a wavelength coverage of 364-800nm. Spectral follow-up of the system was done by the McDonald Observatory's Coud\'e spectrograph \citep{Tull:1994}, mounted on the 2.7m Harlan J. Smith Telescope, on UT 2009-08-08 and 2009-08-31. Observations had spectral resolutions of 40,000 and 60,000; the spectrograph's wavelength range is 380-1000nm.

TRES spectral observations were used to constrain the stellar mass. We obtained archival stellar parameters derived from TRES extracted spectra. Following the Spectral Parameter Classification (SPC) method outlined in \cite{Buchhave:2012}, we obtained best-fit stellar parameters that were then used as priors in our own fitting algorithm. SPC works via a cross correlation between the observed spectrum and a grid of synthetic templates varying in stellar effective temperature, surface gravity, metallicity, and rotational broadening. The cross correlation peak heights in this grid are interpolated via a 4D surface, from which the best fit stellar parameters are determined. SPC uncertainties are estimated via comparison of the spectral parameters determined for a test spectral sample that was also classified via Spectroscopy Made Easy \cite{Valenti:1996}.

\subsubsection{High-resolution direct imaging}
Direct imaging of KOI-134 via adaptive optics (AO) was taken by the Robo-AO laser imaging instrument \citep{Baranec:2013, Baranec:2014} on UT 2012-07-16. The Robo-AO observations were gathered using the Palomar 1.5-m telescope at Palomar Observatory in San Diego, California. It has a 44{\arcsec} {$\times$} 44{\arcsec} field of view and spans 600-950 nm, with a pixel scale of 43.1 mas per pixel \citep{Ziegler:2017}. The images were reduced following the methods described in \cite{Ziegler:2017}. These observations had an \textit{i}'-band {$\Delta$}mag of 3.0 at 3.0\arcsec separation and an estimated PSF of 0.12{\arcsec}, and no stellar companions with spectral types earlier than M4 were detected \cite{Ziegler:2017}. The 5{$\sigma$} contrast curve and direct image are shown in Extended Data Fig. \ref{fig:directimage}. The sensitivity curve rules out any companions with a {$\Delta$}mag of 6 within a minimum separation of $\sim$1\arcsec. Likewise, the direct image shows no sources within 4\arcsec of KOI-134. 

\subsubsection{Stellar parameters}
\label{subsubsection:SEDFitting}
KOI-134 is a F9V-type star with an apparent \textit{Kepler} magnitude of 13.7. The \textit{TESS} Input Catalog (TIC) reports a mass of {\ticMass}{$M_{\odot}$}, a radius of {\ticRadius}{$R_{\odot}$}, and effective temperature of {\ticTeff}\,K for KOI-134.

We conducted preliminary estimates of stellar parameters with {\tt astroARIADNE} (Supplementary Fig. 2; Supplementary Table 2). Stellar fitting was conducted simultaneously during the light curve analysis, the methods for which are laid out in the Methods section. Stellar parameters were fit to isochrones to estimate age in addition to the other derived parameters from the first fitting routine. During the fit, we extrapolated the parameters needed for the isochrones from a spectral energy distribution (SED), which used the following bands: Gaia DR2 {$B_{P}$}, G, and {$R_{P}$}; 2MASS J, H, and {$K_{s}$}; and WISE W1 and W2.

We use the {\tt astroARIADNE} fit results with our transit results to constrain stellar parameters (see the Light Curve Analysis section). However, we caution that directly constraining parameters from the photometry alone can sometimes lead to underestimated errorbars \citep{Tayar:2022}. Therefore, we performed another independent fit following the methodology of \citep{Eastman:2019}, which includes an error scaling term as a free parameter. The error scale ensures that stellar uncertainties were scaled properly when combining stellar and planet likelihoods. This parameter scales the observed photometric errors to better match the model uncertainties \citep{Eastman:2019}. We present the stellar mass and radius from this alternative fit in Extended Data Table \ref{tab:fit}.

\subsubsection{Validation of planetary nature of KOI-134 b}
Multiple independent methods support the planetary nature of KOI-134\,b. From a spectroscopic perspective, the approximate RV semi-amplitude for a 1{$M_\text{Jup}$} planet around a Sun-like star is on the order of several tens to several hundreds of meters per second, which would not be detectable with the TRES instrument. Based on the measured radial velocities from TRES, we derived a 3{$\sigma$} RV semi-amplitude upper limit of 454 m/s, which corresponds to a 3{$\sigma$} mass upper limit of {\bUpperMSinI}{\mjup} for KOI-134\,b. The observed RV signal cannot be caused by another star, since the radial velocity signature of a more massive object would have been detected. Moreover, adaptive optics observations do not reveal the presence of any stellar companions, and the light curve does not show any appreciable secondary eclipses. This indicates that the transit timing variations we see are caused by another planet orbiting KOI-134. These non-detections provide significant evidence for ruling out the eclipsing binary case. Supplementary Fig. 3 depicts the radial velocity observations from TRES that were taken during the 2013 observing season, with a {\tt RadVel} \cite{radvel} model that was generated using our best-fit dynamical parameters. The two observations from the 2012 season were not included, as these contain outliers and we cannot comment on the seasonal variation of the instrument. We do not use the radial velocity signals to constrain the mass of the planet in our TTV models due to the relatively large stellar radial velocity jitter. 

\subsubsection{Light curve analysis}
\label{subsubsection:LC}
We analyzed \textit{Kepler} light curves from each transit epoch to obtain planet parameters for KOI-134\,b. With the {\tt emcee} package \citep{Foreman-Mackey2013}, we use a Markov Chain Monte Carlo (MCMC) to fit a {\tt batman} \citep{Kreidberg(2015)} transit model to the \textit{Kepler} data. SC data was used instead of LC data when available, due to the inherently higher time resolution of the 58.9-second SC data over the 29.4-minute LC data \citep{Murphy:2012}.  Best-fit transit times are reported in Supplementary Table 3. The {\tt batman} model light curves generated with these best-fit transit times are plotted with the \textit{Kepler} data, shown in Fig. \ref{fig:riverplot}.

We fit transit epochs separately, with each {\tt batman} model sharing the same planet and stellar parameters apart from the transit center. Each transit is fit individually, as the transit centers of each epoch are free parameters. To account for the smearing of the transit shape due to the longer integration time of LC data, we super-sample the light curve by a factor of 7. This means that each exposure is split into 7 evenly-spaced samples, then the weighted average value of the light curve is returned, such that the mean exposure time matches that of the observations. Simultaneous stellar fitting of MIST isochrones \citep{MISTModels} was conducted as part of the MCMC.

The following planet parameters were left free: period, {${\rpl}/{\rstar}$}, impact parameter, {\esinw}, {\ecosw}, and transit center for each epoch. We tightly constrained the period in order to constrain the semimajor axis with a tight Gaussian prior centered around approximately 67.59 days and a 1-{$\sigma$} error of 0.2 days. The impact parameter was allowed to range between 0 and 1. Because of the degeneracy between eccentricity ({$e$}) and longitude of periastron ({$\omega$}), we defined {\esinw} and {\ecosw} as free parameters and recovered {$e$} and {$\omega$} after the fit was complete. For both {\esinw} and {\ecosw}, we implemented a loose uniform prior between -0.89 and 0.89 (so that $e <$ 0.8 due to stability constraints). 

Stellar mass, radius, {$T_{\rm{eff}}$}, [Fe/H], age (in Gyr), and limb darkening coefficients  {$q_{1}$} and {$q_{2}$} were included as free parameters. Interstellar extinction, parallax, and distance were all left fixed. The extinction value of 0.17 was based on our {\tt astroARIADNE} fit, which uses the {\tt DUSTMAPS} Python package \citep{Green:2018} to derive a line-of-sight extinction from the SFD Galactic dust map \citep{Schlegel:1998, Schlafly:2011}. Stellar mass, stellar radius, $T_\text{eff}$, and [Fe/H] were constrained with the posteriors from the {\tt astroARIADNE} fit (Supplementary Table 2). Instead of fitting for quadratic limb darkening coefficients {$u_{1}$} and {$u_{2}$}, we fit for limb darkening parameters {$q_{1}$} and {$q_{2}$} and obtained {$u_{1}$} and {$u_{2}$} using equations 15 and 16 of \cite{Kipping:2013}. We defined a uniform prior between 0 and 1 for both {$q_{1}$} and {$q_{2}$}. Fitted limb darkening coefficients are summarized in Table \ref{tab:fit}.

To investigate the system's visible transit duration variations (TDVs), we perform an additional light curve fit using transit duration as a free parameter. The per-epoch transit durations are summarized in Supplementary Table 3. The transit times we obtain from this step are within 1-{$\sigma$} of the transit times derived from the original light curve analysis. Therefore, we still adopt the transit times that that were estimated using our original fit (Supplementary Table 3). In this analysis, we re-parametrize the {\tt batman} model by deriving the impact parameter and inclination from this duration using Eq. 16 from \cite{SeagerMallenOrnelas2003} assuming circular orbits. Other parameters ($t_i$, $P$, {${R_p}/{R_*}$}, $q_1$, $q_2$, $M_*$, $R_*$) remain free but are now constrained with Gaussian priors using the values from our previous light curve fit.

\subsubsection{Transit timing and transit duration dynamic modeling}
\label{subsubsection:ttvmodel}

KOI-134\,b exhibits TTVs with an amplitude of {$\sim$}20 hours, as shown in Fig. \ref{fig:riverplot}. This is due to resonant effects from a non-transiting companion. Through the use of the {\tt REBOUND} package \citep{rebound}, we modeled the system's TTVs and TDVs to find the architecture that would reproduce the signals that we observe. We made use of {\tt REBOUND}'s {\tt IAS15} integrator to model the planets' orbital dynamics (see \cite{ias15} for detailed methods on {\tt IAS15}) to compute the joint TTV and TDV likelihood. With each set of system parameters, we estimate the transit center at each epoch using a bisection method during the n-body integration to determine when the coordinate of the planet changes sign in the sky plane along its orbit. The transit time centers from {\tt REBOUND} are then mapped to the observed transit times by applying a constant offset fitted by least-square method after unit conversion. We also use the recorded orbital elements corresponding to each transit to calculate the transit duration using Eq. 15 of \cite{Kipping:2010} to compare to the observed transit durations.  

We defined 14 free parameters in our $N$-body simulations: planet masses {$m_{1}$} and {$m_{2}$}, orbital periods {$p_{1}$} and {$p_{2}$}, arguments of periastron {$\omega_{1}$} and {$\omega_{2}$}, eccentricities {$e_{1}$} and {$e_{2}$}, longitudes of the ascending node {$\Omega_{1}$} and {$\Omega_{2}$}, the inclination of the non-transiting planet {$i_{2}$}, mean anomalies {$M_{1}$} and {$M_{2}$}, and the impact parameter {$b_1$}. The stellar mass and radius were fixed, taken from the best fit parameters generated by the isochrone fitting during light curve analysis. We performed a grid search to find optimal starting points for our free parameters, and initialized our MCMC at these results (Supplementary Fig. 4).

Out of all the 8 mean motion resonances we explored, we only found one set of stable solutions near the inner 2:1 resonance of KOI-134 b. We found two solutions that fit both the TTV and TDV data equally well. One solution indicates that KOI-134\,c is a sub-Saturn (0.21 $M_\text{Jup}$) with a mutual inclination of $\approx 15^{\circ}$ relative to KOI-134\,b; the other solution indicates that KOI-134\,c has a mass of 0.7 \mjup\, and a mutual inclination of $\approx 50^{\circ}$ relative to KOI-134\,b. Further analysis shows only the former solution exhibits long-term stability (see the section on dynamic modeling) and is presented in Table \ref{tab:nbody}. 

We explore the mass-eccentricity posterior space of the KOI-134 system further by allowing eccentricity of both planets to be as high as 0.8 while exploring the posterior spaces, and find no solutions with reduced $\chi^2$ better than 2.14 in the high eccentricity region. In contrast, our reduced $\chi^2$ for the low eccentricity region was 1.52.

\subsubsection{Photodynamical modeling}
As a check on the robustness of our parameters, we conduct photodynamical modeling with {\tt PyTTV} \cite{Korth:2020} to jointly fit \textit{Kepler} photometry and $N$-body simulations. We follow the methodology of \cite{Korth:2023} for this analysis, which entails simultaneous modeling of transit light curves and TTVs. Instead of producing only mid-transit times through $N$-body simulations, these simulations now produce full light curves. This means that instead of employing the usual approach of only comparing transit times, this model takes into account features such as transit shape to compare the dynamically-simulated light curve with \textit{Kepler} observations and precisely determine planet parameters. From this approach, we find results that are consistent with our non-photodynamic TTV analysis (Table \ref{tab:nbody}). 

\subsubsection{Resonant dynamics}
\label{subsubsection:stability}

We use {\tt REBOUND} to further analyze the system stability and evolution. We draw from the posterior near both sets of solutions found in our TTV models and integrate the system using the ``ias15" integrator to 10 Myrs. Here we define to two sets of solutions: the ``modest mutual inclination solution ($\approx 15^{\circ}$)" and ``the high mutual inclination solution ($\approx 50^{\circ}$)". The ``modest mutual inclination" solution is always dynamically stable to the end of our integration. However, the ``high mutual inclination" solution often led to the ejection of KOI-134\,c in less than $10^4$ years. We further examine the evolution of the resonance angles for the two sets of solutions. The ``modest mutual inclination" solution shows clear libration around the eccentricity resonance angle $\Theta_{e,c}$, while the ``high mutual inclination" solution does not show any libration for neither the eccentricity resonance angles, nor the inclination resonance angles. In addition, we examine the position of the two solutions in the X-$\Gamma$' resonance domain. The ``modest mutual inclination" solution is firmly in the resonance, while the ``high mutual inclination" solution is just narrow of the 2:1 resonance. We conclude that only the ``modest mutual inclination" solution is feasible to reproduce the orbital characteristics of KOI-134 system.   

The best-fit ``modest mutual inclination" solution indicates that the KOI-134 system has a TTV amplitude of $\approx$ 20 hours. We found that the amplitude of the libration angle {$\Theta_{e,c}$} is 70{$^{\circ}$}. In our integrated solutions, we find that the the resonant libration timescale - i.e., the timescale for the biggest variations in both planets’ semi-major axes and the inner planet’s eccentricity - is $\sim$4.5 years. The libration time scale represents the TTV time scale in the cases where planets are librating in resonance \citep{NesvornyVokrouhlicky:2016}. The outer planet’s eccentricity varies on apsidal alignment timescale ($\sim$4 years). 

Near an orbital resonance, the eccentricity of a planet can be broken into free and forced components; the forced eccentricity of a planet is governed by how close the companion is to resonance, while the free eccentricity is independent of the resonance and has an arbitrary phase and magnitude. Therefore, the changes in eccentricity stem from variations in the forced component due to oscillations around the 2:1 resonance. Additionally, free eccentricities can affect the amplitude and phase of TTVs, creating a degeneracy with planet mass. Following \cite{Dawson:2021}, we estimate that KOI-134\,b has a forced eccentricity of 0.206 and a free eccentricity of 0.034. KOI-134\,c has forced and free eccentricities of 0.197 and 0.196, respectively. These nonzero free eccentricities could indicate excitation via planetary interactions, if free eccentricity is generally expected to be damped away via resonant repulsion \cite{LithwickWu:2012}. 

\subsubsection{Possible mutual inclination excitation via disk migration}

We investigate one possible origin of the KOI-134 system through disk migration using a time symmetric Hermite integrator \citep{Aarseth:2003}.  We assume that the inner planet has already arrived at the inner edge of the protoplanetary disk (so it no longer interacts with the disk) at the start of the simulations. The inner planet remains at its original location until the outer planet has migrated near it. The interactions between the outer planet and the disk, i.e. the effect of disk migration, eccentricity damping, and inclination damping are approximated by their corresponding time scales ($\tau \rm_{m}$, $\tau \rm_e$, and $\tau \rm_{i}$). The total acceleration acting on the planet is expressed as 
\begin{eqnarray}
\label{force}
\frac{d}{dt}\textbf{V}_i =
 -\frac{G(M_*+m_i)}{{r_i}^2}\left(\frac{\textbf{r}_i}{r_i}\right)
+\sum _{j\neq i}^N Gm_j \left[\frac{(\textbf{r}_j-\textbf{r}_i
)}{|\textbf{r}_j-\textbf{r}_i|^3}- \frac{\textbf{r}_j}{r_j^3}\right]
\nonumber\\
- 2{ ({\bf v \cdot r}) {\bf r}  \over r \tau \rm_e}
 - 2{ ({\bf v \cdot k}) {\bf k}  \over \tau \rm_ {i}}- { {\bf v}  \over  \tau_{\rm m}},~~~~~~~~~~~~~~~ \label{eq:eqf}
\end{eqnarray}
where $M_*$ and $m_i$ represent the mass of the central star and the $i$th planet. 

We perform 27 $N$-body simulations with different sets of damping timescales, migration timescales, and disk depletion timescales for the outer planet. We assume that the outer planet migrates with a timescale in the range of ${\tau_m}\in[10^5, 10^6]$ yrs, the eccentricity damping timescale is [$10^{-3}, 10$]{$\times$}$\tau_m$, and the inclination damping timescale is comparable to the eccentricity damping timescale.  

We find that migration can easily reproduce a 2:1 eccentricity type resonance, with the two planets then convergently migrating inward to their current positions. With $\tau_e/\tau_m<0.1$, the eccentricities of the inner and outer planets do not exceed 0.5 and 0.15, respectively. This only produces low mutual inclinations ($< 5^{\circ}$). For the cases with $\tau_e/\tau_m>0.1$, the mutual inclination can be excited to larger than 10{$^{\circ}$} as the two planets encounter the 4:2 inclination type resonance. However, triggering the inclination resonance requires eccentricity excitation much larger than what we observe ({$e > 0.5$} in the case of the inner planet). Eccentricity damping does not change this feature. Therefore, if the mutual inclination of the KOI-134 system is obtained through orbital migration in the protoplanetary disk, then there needs to be an explanation for the reduction of the eccentricity to the observed values.

Beyond MMR capture through disk migration, other mechanisms have been proposed to excite mutual inclination. For instance, as the host star rotates slower due to magnetic braking, its quadrupole moment becomes reduced and this can lead to a decrease in the nodal precession rate of the close-in planets. The decrease in the nodal precession rate may cause resonances between the planets and lead to a larger mutual inclination \citep[e.g.,][]{Faridani23, Faridani24}. However, KOI-134\,c is not sufficiently close for stellar $J_2$ to play a significant role. Moreover, a protoplanetary disk may also lead to nodal precession and the dispersal of the disk can also cause sweeping resonances and inclination/eccentricity excitation \citep[e.g.,][]{Nagasawa03, Petrovich20}. Nevertheless, massive disks are needed in order to drive a fast nodal precession and resonant inclination excitation. Therefore, it remains unclear what combination of mechanisms created the unique architecture of the KOI-134 system, and further theoretical study is needed to constrain its formation history.

\subsubsection{Search for additional transits in \textit{TESS}}
Eight total sectors of \textit{TESS} observations (Sector 14, 15, 40, 41, 54, 55, 74 and 75) were taken of KOI-134. Due to the relative faintness of the star in the \textit{TESS} band ($T = $13.2), the per-point scatter of the \textit{TESS} Quick-Look Pipeline light curves \cite{Huang:2020b} are comparable to the transit depth of KOI-134\,b. The extremely large amplitude TTVs make the standard Box Least Square \cite{Kovacs:2002, Hartman:2012} search of the planet transits in the \textit{TESS} light curves impossible. We draw 160 realizations from the the posteriors of KOI-134 b's orbital parameters, and propagate them to the time frame of the \textit{TESS} observations. We found that three separate transits of KOI-134 b are predicted to be observed in Sector 14, Sector 41, and Sector 54 (Supplementary Fig. 5). We visually examined these segments of light curves and found two candidate transit events in Sector 14 and Sector 54. The predicted transit in Sector 41 may have fallen into the TESS data down-link gap between orbit 89-90. Upon further investigating the two candidate events, we found that the event in Sector 14 coincides with a \textit{TESS} momentum dump event (see TESS Data Release Note 19). Future high signal to noise transit observations of KOI-134\,b are needed to help test whether a true transit event was observed in \textit{TESS} sector 54.

\subsection*{Data Availability}
The \textit{Kepler} data used in this analysis can be accessed via the \textit{Kepler} Data Search \& Retrieval Tool (\url{https://archive.stsci.edu/kepler/data_search/search.php}). Any other datasets that were generated can be obtained from E. N. upon reasonable request.

\subsection*{Code Availability}
This study makes use of the following publicly-available packages: {\tt astroARIADNE}, {\tt batman}, {\tt numpy}, {\tt matplotlib}, {\tt RadVel}, {\tt REBOUND}, {\tt scipy}. The scripts used for this analysis are located at \url{https://github.com/enabbie/KOI134}.

\section*{Extended Data}
\setcounter{figure}{0}
\setcounter{table}{0}

\begin{table}[htp]
\caption{
  Stellar and Planet Parameters for KOI-134
  \label{tab:fit}
}

\begin{tabular}{lcr}

\toprule
Parameter & Value & Source \\
\midrule

\noalign{\vskip -3pt}
\textbf{Catalog Information} & & \\
~~~~R.A. (h:m:s)                      & \starRA     & Gaia DR3\\
~~~~Dec. (d:m:s)                      & \starDec    & Gaia DR3\\
~~~~Epoch							  & \starRefEpoch   & Gaia DR3 \\
~~~~Parallax (mas)                    & \starParallax  & Gaia DR3\\
~~~~$\mu_{ra}$ (mas yr$^{-1}$)        & \starPMRA   & Gaia DR3 \\
~~~~$\mu_{dec}$ (mas yr$^{-1}$)       & \starPMDec & Gaia DR3\\
~~~~Gaia DR3 ID                       & \starGaiaID   &  \\
~~~~KIC ID                            & \starKICID & \\
~~~~KOI ID                            & \starKOIID  & \\
~~~~TIC ID                            & \starTICID & \\
\textbf{Photometric properties} & & \\
~~~~$Kepler$ (mag)\dotfill            & \starKepMag & KIC         \\
~~~~$TESS$ (mag)\dotfill            & \starTMag  & TIC v8.2         \\
~~~~$Gaia$ (mag)\dotfill            & \starGaiaMag & Gaia DR3     \\
~~~~Gaia RP (mag)\dotfill          & \starGaiaRPMag & Gaia DR3                 \\
~~~~Gaia BP (mag)\dotfill          & \starGaiaBPMag & Gaia DR3                 \\
~~~~$V_J$ (mag)\dotfill             & \starVMag & APASS DR10      \\
~~~~$B_J$ (mag)\dotfill             & \starBMag & APASS DR10      \\
~~~~$J$ (mag)\dotfill               & \starJMag & 2MASS           \\
~~~~$H$ (mag)\dotfill               & \starHMag & 2MASS           \\
~~~~$K_s$ (mag)\dotfill             & \starKMag & 2MASS           \\
\textbf{Derived properties} & & \\
~~~~$\mstar$ ($M_{\odot}$)\dotfill      &  $1.407 \pm 0.043$ &  \\
~~~~$\rstar$ ($R_{\odot}$)\dotfill      & $1.667 \pm 0.024$ &        \\
~~~~$\mstar$ ($M_{\odot}$) [alt]\dotfill      &  \starMass & \footnotemark[1]\\
~~~~$\rstar$ ($R_{\odot}$) [alt]\dotfill      & \starRadius & \footnotemark[1]\\
~~~~$\loggstar$ (cgs)\dotfill       & \starLogg &  empirical relation + LC       \\
~~~~$L_{\star}$ ($L_{\odot}$)\dotfill      & \starLuminosity  &    \\
~~~~$\teffstar$ (K)\dotfill        &  \starTeff  & \\
~~~~[Fe/H]\dotfill        &  $0.427 \pm 0.069$  & \\
~~~~$M_V$ (mag)\dotfill &  $14.39\pm$0.02  & Parallax         \\
~~~~$M_K$ (mag)\dotfill &  $8.272\pm$0.015  & Parallax         \\
~~~~Distance (kpc)\dotfill           & \starDistance  & Parallax\\
~~~~\rhostar (\gcmc)\dotfill &  \starRho & empirical relation + LC \\
~~~~$v\,\text{sin}(i)$ (km s$^{-1}$)\dotfill &  $10.59 \pm 0.38$ & \\

\textbf{Limb-darkening coefficients} & & \\
~~~$u_1,Kepler$               \dotfill    & \starKepleruOne     &  \\
~~~$u_2,Kepler$               \dotfill    &  \starKepleruTwo    & \\
[1.5ex]
\textbf{Light curve parameters} & \textbf{KOI-134\,b} \\
~~~$P$ (days)  \dotfill    &  {\bPeriod} {\footnotemark[2]} \\
~~~$T_{14}$ (hr) \dotfill    &   {\bDuration} {\footnotemark[3]} \\
~~~$T_{12} = T_{34}$ (min)   \dotfill    &  \bIngressDuration \\
~~~$\arstar$              \dotfill & \bAOR \\
~~~$\rpl/\rstar$          \dotfill & \bROR \\
~~~$b \equiv a \cos i/\rstar$ \dotfill    & \bImpactParameter \\
~~~$i$ (deg) \dotfill &  \bInclination \\
\textbf{Planetary parameters} & & \\
~~~$\rpl$ ($R_J$) \dotfill &   \bRadius \\
~~~$a$ (AU) \dotfill & \bSemimajorAxis \\
~~~$T_{\rm eq}$ (K) \dotfill &   \bTeq \\
~~~$\langle F \rangle$ ($S_{\oplus}$) \dotfill    & \bIrr & \\
\botrule
\end{tabular}
\footnotetext[1]{We report stellar mass and radius from an independent fit with scaled errors to account for stellar model uncertainties. Please see the Methods section for more details.}
\footnotetext[2]{This orbital period is derived from a linear ephemeris. We do not provide uncertainties on this parameter, as it is not modeled in our light curve analysis.}
\footnotetext[3]{Due to the TDVs, this $T_{14}$ value represents the average transit duration.}
\end{table}

\begin{table}[h]
\caption{
  High-Precision System Parameters for $N$-body Integration
  \label{tab:nbodyhp}
}
\begin{tabular}{lcc}
\toprule
Parameter\footnotemark[1] & KOI-134\,b & KOI-134\,c \\
\midrule
Mass [{\mjup}] & 0.980205 & 0.213663\\
Period [days] & 67.128116 & 33.937346\\
Eccentricity & 0.200958 & 0.285108\\
Argument of of periastron, {$\omega$} [$^\circ$] &  -73.333735 & 90.000000\\
Longitude of ascending node, {$\Omega$} [$^\circ$] & 34.504440 & 223.412554\\
Inclination [$^\circ$] & 89.618729 & 74.509492\\
Mean anomaly, {$M$} [$^\circ$] & 3.830639 & -7.896443\\
\botrule
\end{tabular}
\footnotetext[1]{Because our posterior space is highly asymmetric, we report the mode values instead of median values.}
\end{table}

\begin{figure*}[htp]
\centering
\includegraphics[width=1\textwidth]{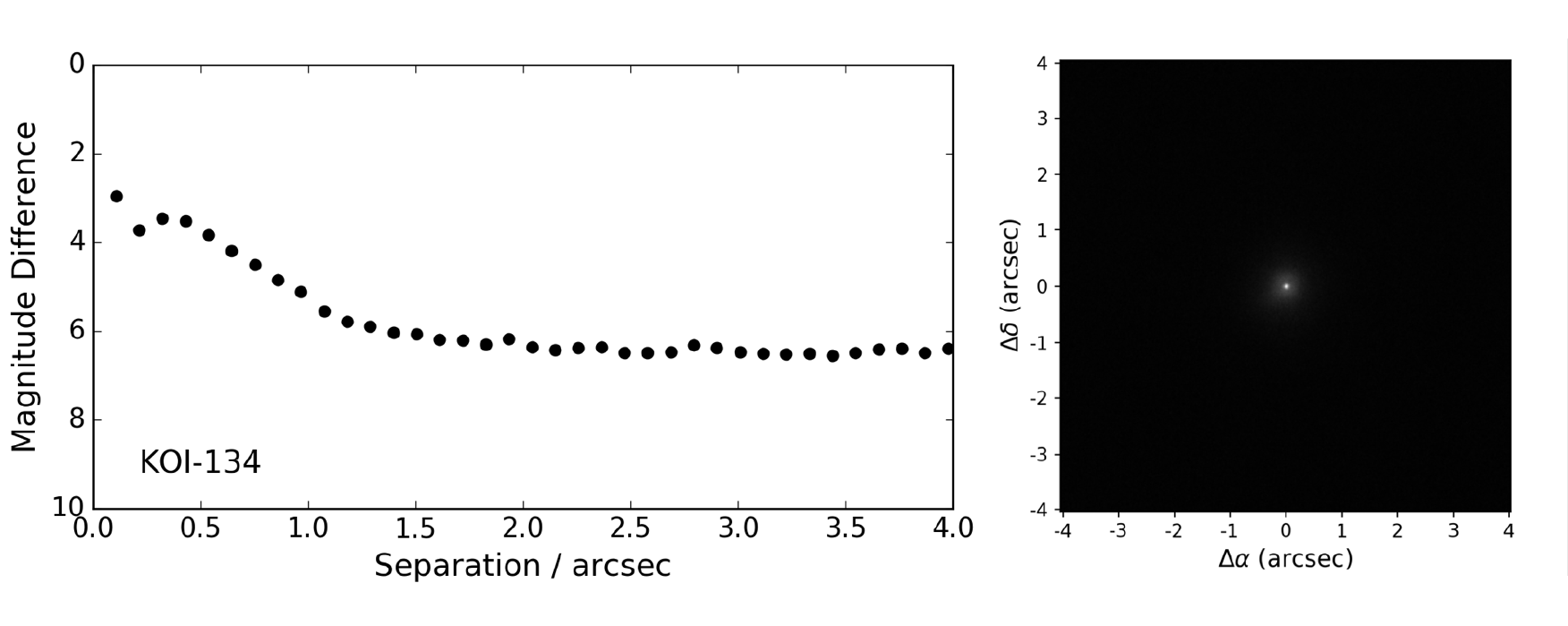}\hfill

\caption{\textbf{Direct imaging observations of KOI-134.} {\textit{Left}: 5{$\sigma$} sensitivity curve depicting the difference in magnitude versus orbital separation. \textit{Right}: High resolution i-band direct imaging of KOI-134 observed with the P60 telescope's Robo-AO instrument. The 8" square image cutout is centered on the star, with a pixel scale of 0.021" per pixel.}}
\label{fig:directimage}

\end{figure*}

\begin{figure*}[htp]
\centering
\includegraphics[width=1\textwidth]{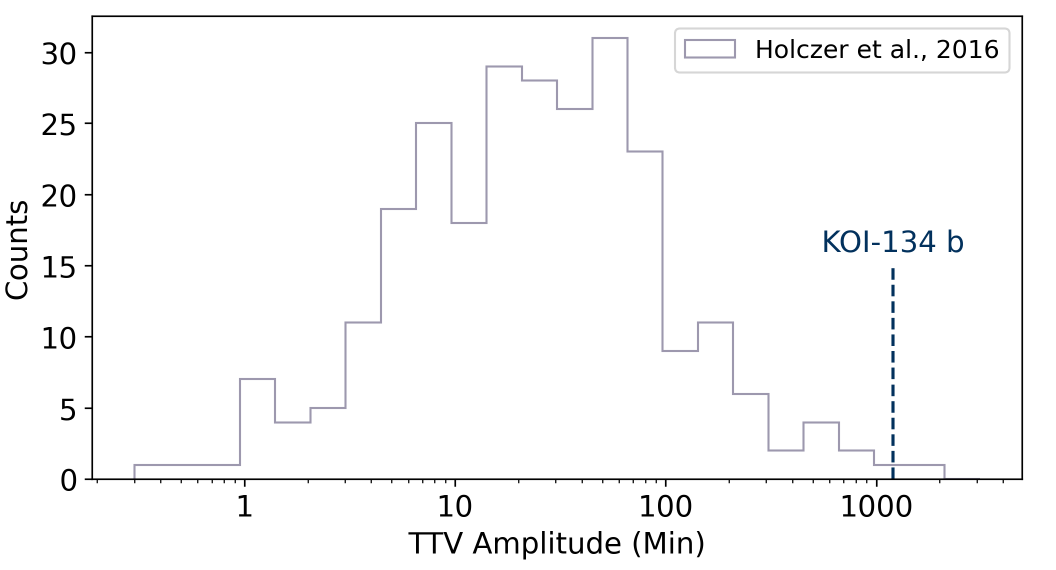}\hfill

\caption{\textbf{KOI-134\,b in context with other Kepler TTV systems.} The histogram shows the distribution of TTV amplitudes from Kepler systems with significant TTVs, taken from Table 5 of the \textit{Kepler} TTV catalog by \cite{Holczer:2016}. The dashed line shows the position of KOI-134\,b in comparison to this population.}
\label{fig:ttvdist}
\end{figure*}

\begin{figure*}[htp]
\centering
\includegraphics[width=1\textwidth]{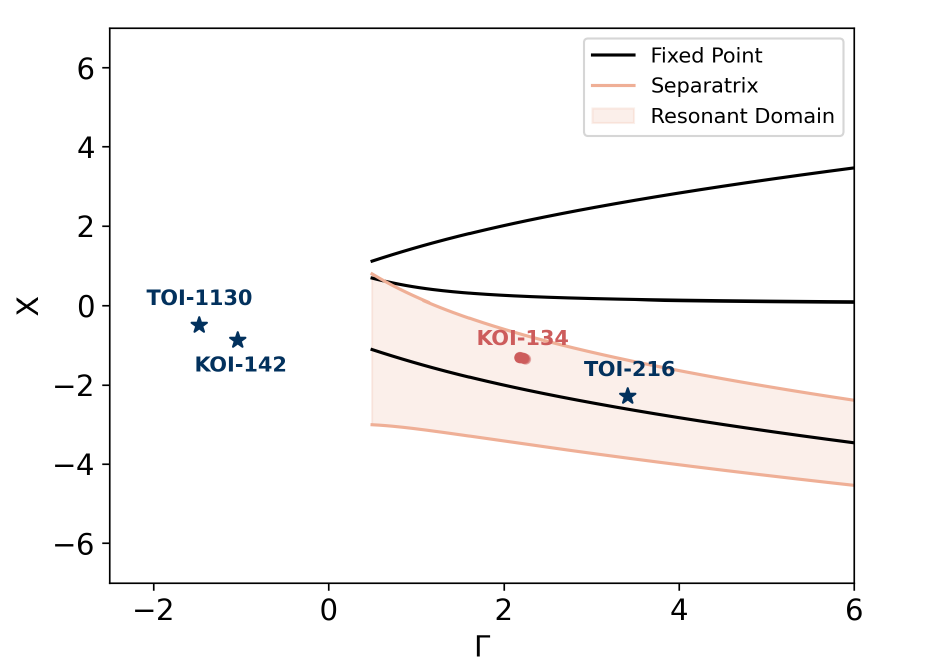}\hfill

\caption{\textbf{The proximity of KOI-134 to an exact 2:1 resonance.} The position of KOI-134 within the 2:1 resonant domain, re-parametrized to a 1-degree of freedom model \cite{Deck:2013}. Randomly-sampled points from the planet parameter posterior distribution are shown in red circles. {$X$} measures the system's separatrices and stable/unstable fixed points, and {$\Gamma$} quantifies how close the system is to an exact resonance. The shaded region denotes the formally-defined resonant domain.}
\label{fig:resonance}
\end{figure*}

\begin{figure*}[htp]
\centering
\includegraphics[width=1\textwidth]{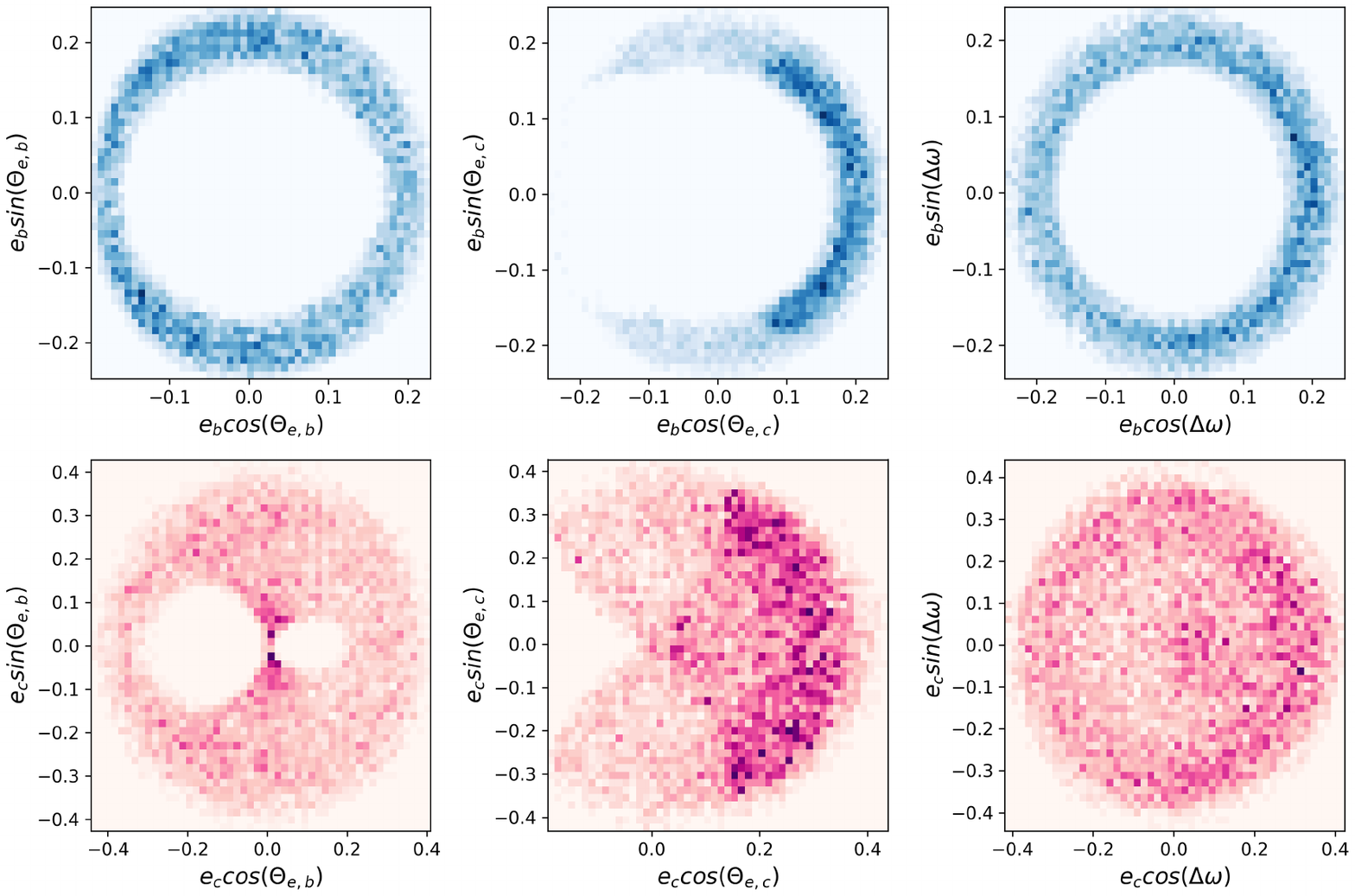}\hfill

\caption{\textbf{Trajectories of the resonant angles of KOI-134\,b and c}. Trajectories of the resonant angles ({${\Theta}_{e,b}$ and ${\Theta}_{e,c}$}) and the secular apsidal angle ({${\Delta}{\omega}$}), as they evolve over a time scale of 1 million years. The color scheme is shared with Fig. \ref{fig:evolution}.}
\label{fig:trajectory}
\end{figure*}

\begin{figure}[htp]
\centering
\includegraphics[width=.5\textwidth]{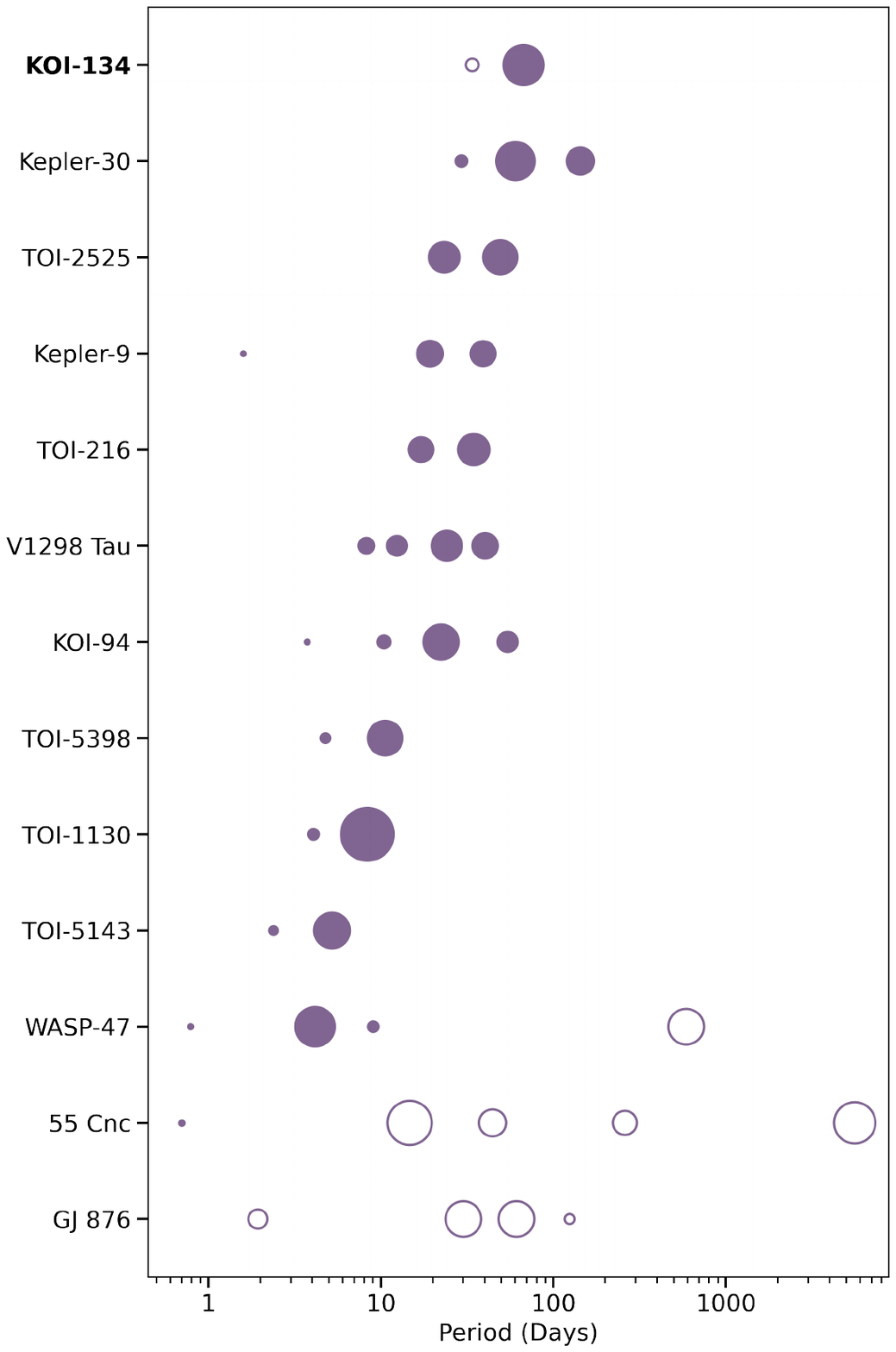}\hfill

\caption{\textbf{The population of near-MMR multi-planet systems with giant planets.} Confirmed multi-planet systems with a giant planet, where at least one pair of orbital periods are near a first-order resonance. The empty circles represent non-transiting planets. Marker size corresponds to relative planet radius.}
\label{fig:population}
\end{figure}

\section*{Declarations}

\subsection*{Correspondence and Request for Materials}
All correspondences and requests for materials should be directed to Emma Nabbie (Email: Emma.Nabbie@usq.edu.au).

\subsection*{Acknowledgments}
We would like to thank R. Dawson for helpful discussions throughout our analysis.
E.N. acknowledges the PhD scholarship provided by the ARC Discovery Project DP220100365.
C.X.H. and A.V. thank the support of the ARC DECRA project DE200101840.
A.V. would like to thank M. Omohundro, K. Deck, A. Vanderburg, and J. Becker for helpful comments during the inception of this study.
G.Z. thanks the support of the Australian Research Council project FT230100517. 
The authors acknowledge support from the Swiss NCCR PlanetS and the Swiss National Science Foundation. This work has been carried out within the framework of the NCCR PlanetS supported by the Swiss National Science Foundation under grants 51NF40182901 and 51NF40205606. J.K. acknowledges support from the Swedish Research Council (Project Grant 2017-04945 and 2022-04043) and of the Swiss National Science Foundation under grant number TMSGI2\_211697.
H.P. acknowledges support by the Spanish Ministry of Science and
Innovation with the Ramon y Cajal fellowship number RYC2021-031798-I. Funding from the University of La Laguna and the Spanish Ministry of Universities is acknowledged.

\subsection{Authors' Contributions}
E.N. led the light curve and $N$-body analyses, interpretation of the results, and preparation of the manuscript. C.X.H. supervised the project, conducted $N$-body simulations, facilitated analysis of the system's TDVs, and contributed to the writing of the manuscript. J.K. and H.P. performed the photodynamical analyses. S.W. conducted disk migration simulations to investigate the system's formation and evolution. A.V. identified the system and contributed to the analysis of the \textit{Kepler} observations and mid-transit times. R.W. supervised the project and contributed to the analysis. G.L. contributed to the theoretical dynamical analyses. D.N.C.L. contributed to analysis and interpretation of disk migration simulations. A.B., D.W.L., and G.Z. were responsible for the TRES radial velocity observations and data reduction.

\subsection{Competing Interests}
The authors declare no competing interests.

\newpage
\renewcommand{\figurename}{Supplementary Figure}
\renewcommand{\tablename}{Supplementary Table}
\setcounter{figure}{0}
\setcounter{table}{0}
\section{Supplementary Text}
\subsection{Constraining Stellar Parameters with ARIADNE}
To obtain more precise stellar characterization, we constructed a spectral energy distribution (SED) from the star's broadband photometry. Our SED, shown in Supplementary Fig. \ref{fig:SED}, consisted of the following photometric bands: Pan-STARRS1 gp1, rp1, ip1, zp1, and yp1 \citep{PS1}; Gaia DR2 {$B_{P}$}, G, and {$R_{P}$} \citep{GaiaDR3}; 2MASS J, H, and {$K_{s}$} \citep{Skrutskie2006}; and WISE W1 and W2 \citep{Wright2010}. We implemented a magnitude uncertainty floor of 0.03 to avoid over-constraining the fit. With the Python package {\tt astroARIADNE} \citep{ariadne}, we fit the SED to multiple stellar atmosphere models and used a Bayesian Model Averaging (BMA) method to find the best fit parameters for {$T_\text{eff}$}, radius, {log$g$}, and {[Fe/H]}. The BMA fit calculated the weighted average for the best-fit parameters generated by the PHOENIXv2 \citep{Husser2013}, BTSettl-AGSS2009 \citep{Allard2011}, Kurucz \citep{KuruczModel}, and ATLAS9 \citep{Castelli2004} models to account for any systematic biases. Gaussian priors for {$T_\text{eff}$} and {[Fe/H]} were defined from TRES spectroscopy; we initialized {$T_\text{eff}$} at 6140K with an error of 125K, and {[Fe/H]} at .48 dex with an error of .08 dex. A prior on the distance was obtained using the parallax from Gaia DR3 \citep{Gaia,GaiaDR3}. We corrected for the Gaia DR3 parallax offset following \cite{DR3ParallaxCorrection}. The output values from {\tt astroARIADNE} (Supplementary Table \ref{tab:ariadne}) then served as inputs for further isochrone fitting and light curve analysis.

\subsection{Grid search of transit timing MCMC parameters}
Because of the complex posterior space, we used a grid search technique to start our MCMC at the most optimal points. We performed a linear fit of the calculated transit times to obtain the period and residual and initialize the grid search. We generated a fine period ratio grid to explore eight resonances (1:2, 2:3, 3:4, 1:3, 4:3, 3:2, 2:1, 3:1). The individual resonance grids were comprised of 55 points each, 0.1 units wide and centered around the resonant period ratio. The likelihood values for each period ratio are summarized in Supplementary Fig. \ref{fig:pratio}. Coarser grids were made for log mass, {$\omega$}, {$e$}, and {$\Omega$} of each planet. The boundaries for these grids were as follows: 5 {$M_{\oplus}$} {$\le$} {$m_{1}$}, {$m_{2}$} {$\le$} 10 {$M_\text{Jup}$}; 0{$^{\circ}$} {$\le$} {$\omega_{1}$} {$\le$} 320{$^{\circ}$}; 0{$^{\circ}$} {$\le$} {$\omega_{2}$} {$\le$} 240{$^{\circ}$}; 0.1 {$\le$} {$e_{1}$}, {$e_{2}$} {$\le$} 0.5; 0{$^{\circ}$} {$\le$} {$\Omega_{1}$}, {$\Omega_{2}$} {$\le$} 180{$^{\circ}$}.

For the parameters that were not explored in the grid search, we manually determined their starting positions. We used a log uniform prior on the mass of each planet, with the mass of both planets was bounded between 5{$M_{\oplus}$} and 10 {$M_\text{Jup}$}. In addition, $e_1$ is also limited to be smaller than 0.6 to enhance the possibility of long term stability. The period ratios were limited to values within 10\% of each resonance that was explored during the grid search. Our MCMC probed the parameter space around each resonance individually to reveal any potential degeneracies in the period ratio.

The remaining free parameters were all bounded using uniform box priors. Where applicable, the boundaries on parameters were extrapolated from the results of the transit fitting. Comprehensively, the boundaries for the uniform priors were as follows: 0{$^{\circ}$} {$\le$} {$\omega_{1}$}, {$\omega_{2}$} {$\le$} 360{$^{\circ}$}; 0 {$\le$} {$e_{1}$}, {$e_{2}$} {$\le$} 0.6; 0{$^{\circ}$} {$\le$} {$\Omega_{1}$}, {$\Omega_{2}$} {$\le$} 180{$^{\circ}$}; 0{$^{\circ}$} {$\le$} {$i_{2}$} {$\le$} 90{$^{\circ}$}; -180{$^{\circ}$} {$\le$} {$M_{2}$}, {$M_{2}$} {$\le$} 180{$^{\circ}$}. 

\section{Supplementary Figures and Tables}
\begin{figure*}[htp]
\centering
\includegraphics[width=1\textwidth]{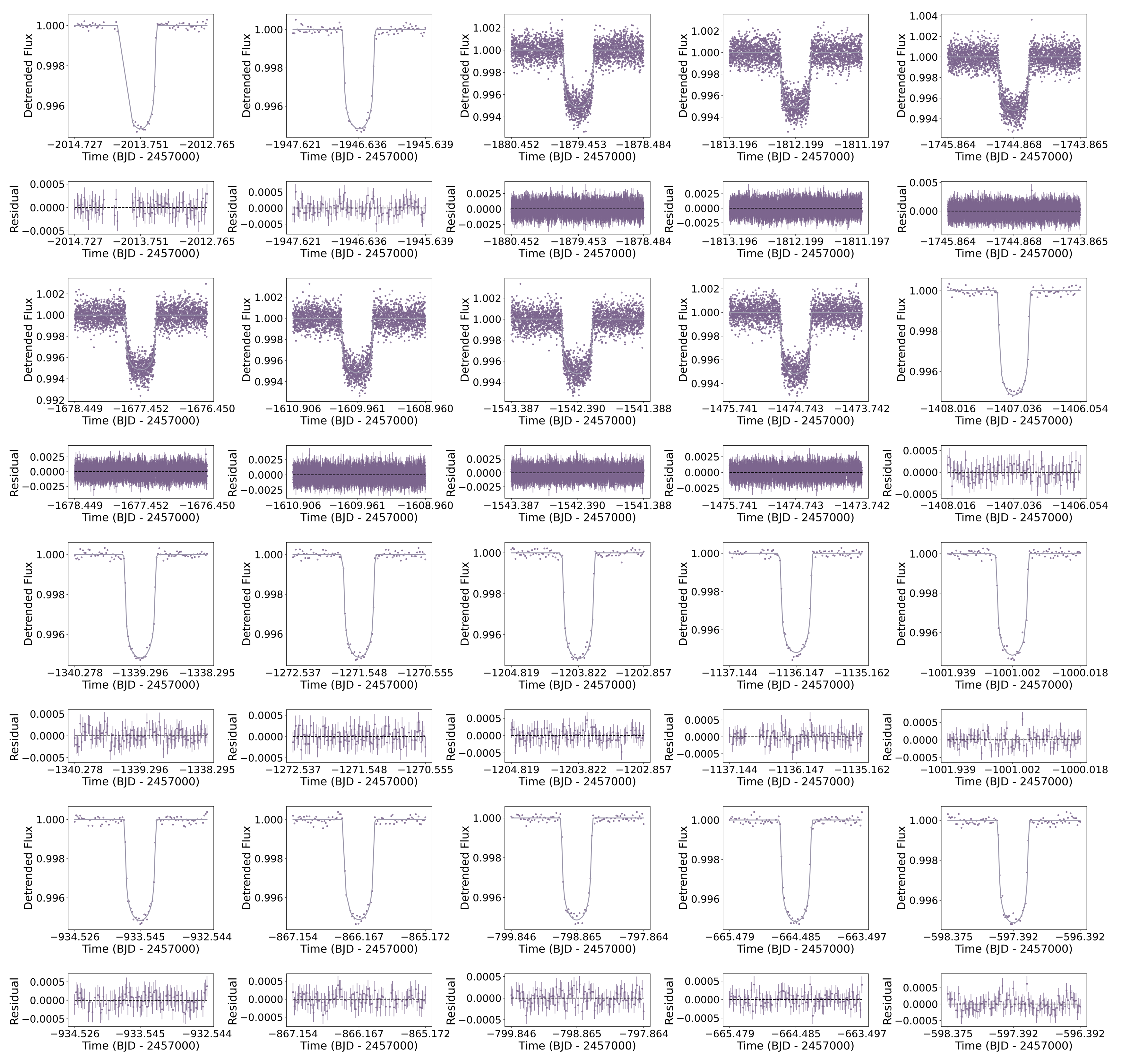}\hfill

\caption{\textbf{Individual transits of KOI-134\,b.} Multi-epoch light curves showing transits of KOI-134\,b. The best-fit {\tt batman} model with variable transit duration for each epoch is overplotted in grey. The residuals are shown in the panels directly below the transit light curves for each of the 20 epochs of \textit{Kepler} data, with errorbars based on the mean absolute deviation of the light curve flux.}
\label{fig:lightcurves}

\end{figure*}

\begin{figure}[htp]
\centering
\includegraphics[width=.5\textwidth]{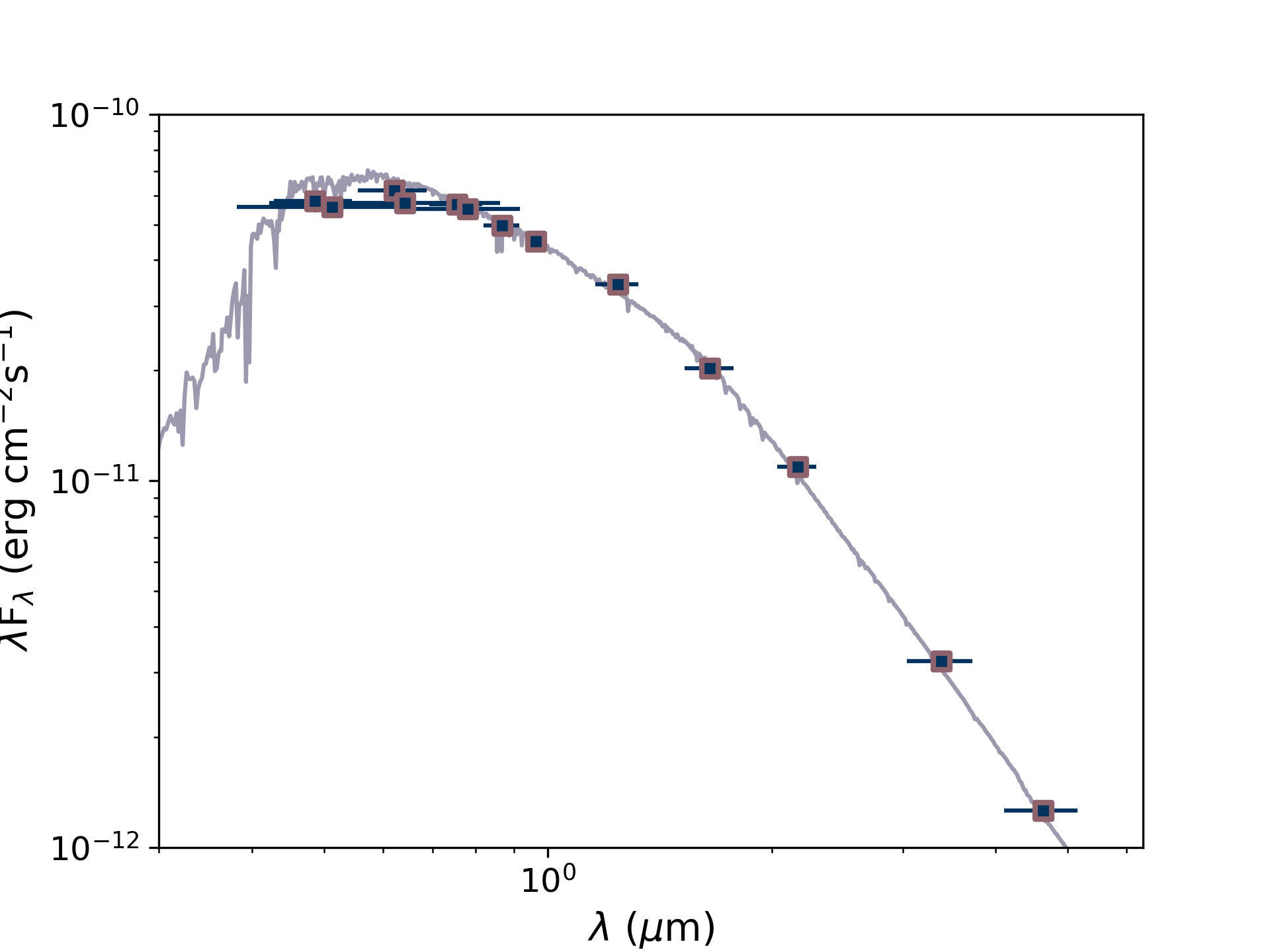}\hfill

\caption{{\textbf{Spectral energy distribution of KOI-134.} Stellar spectral energy distribution with an ATLAS9 model overlaid. The individual points reflect the 13 archival photometric bands taken from the \textit{Gaia} DR2 Crossmatch Catalog. Bandpass widths for each band are represented by the horizontal errors.}}
\label{fig:SED}

\end{figure}

\begin{table}[h]
\caption{SG2 Reconnaissance Spectroscopy}\label{tab:recon}
\begin{tabular}{p{0.1\linewidth}p{0.1\linewidth}p{0.1\linewidth}p{0.1\linewidth}p{0.08\linewidth}p{0.15\linewidth}p{0.1\linewidth}p{0.1\linewidth}p{0.05\linewidth}p{0.05\linewidth}}
\toprule
Telescope & Instrument & Date (UT) & Exposure Time & SNR & BJD & $RV_{CCF}$ ms$^{-1}$ & $\sigma_{CCF}$ & $RV_{LSD}$ ms$^{-1}$ & $\sigma_{LSD}$ \\
\midrule
FLWO (1.5 m) & TRES & 2012-11-07 & 3000 & 26.8 & 2456238.608781 & 108 & 74 & -119 & 48\\
FLWO (1.5 m) & TRES & 2012-11-27 & 3000 & 30.2 & 2456258.619465 & 478 & 77 & 161 & 42\\
FLWO (1.5 m) & TRES & 2013-04-04 & 2700 & 27.9 & 2456386.931045 & 210 & 70 & -21 & 110\\
FLWO (1.5 m) & TRES & 2013-04-08 & 2900 & 25.5 & 2456390.968988 & 188 & 97 & 263 & 72\\
FLWO (1.5 m) & TRES & 2013-04-21 & 2700 & 24.4 & 2456403.949126 & 34 & 77 & 209 & 51\\
FLWO (1.5 m) & TRES & 2013-05-30 & 2300 & 25.8 & 2456442.871811 & 182 & 81 & 266 & 62\\
FLWO (1.5 m) & TRES & 2013-06-03 & 2700 & 27.5 & 2456446.815183 & 220 & 88 & 133 & 69\\
FLWO (1.5 m) & TRES & 2013-06-25 & 2700 & 25.3 & 2456468.755581 & 143 & 120 & 151 & 90\\
FLWO (1.5 m) & TRES & 2013-07-30 & 2700 & 28.4 & 2456503.778636 & 0 & 81 & 0 & 56\\
\botrule
\end{tabular}
\end{table}

\begin{table}[h]
\caption{
  SED Fit Results
  \label{tab:ariadne}
}
\begin{tabular}{cccccccccc}
\toprule
Parameter & Value \\
\midrule
{$T_\text{eff}$ [K]} & {\ensuremath{6153^{+61}_{-129}}}\\
Log(g) [dex] & {\ensuremath{4.134^{+0.029}_{-0.029}}}\\
{[Fe/H] [dex]} & {\ensuremath{0.434^{+0.076}_{-0.073}}}\\
Radius [$R_{\odot}$] & {\ensuremath{1.671^{+0.029}_{-0.031}}}\\
$A_V$ &  $0.17^{+0.084}_{-0.087}$\\
\botrule
\end{tabular}
\end{table}

\begin{figure}[htp]
\centering
\includegraphics[width=.5\textwidth]{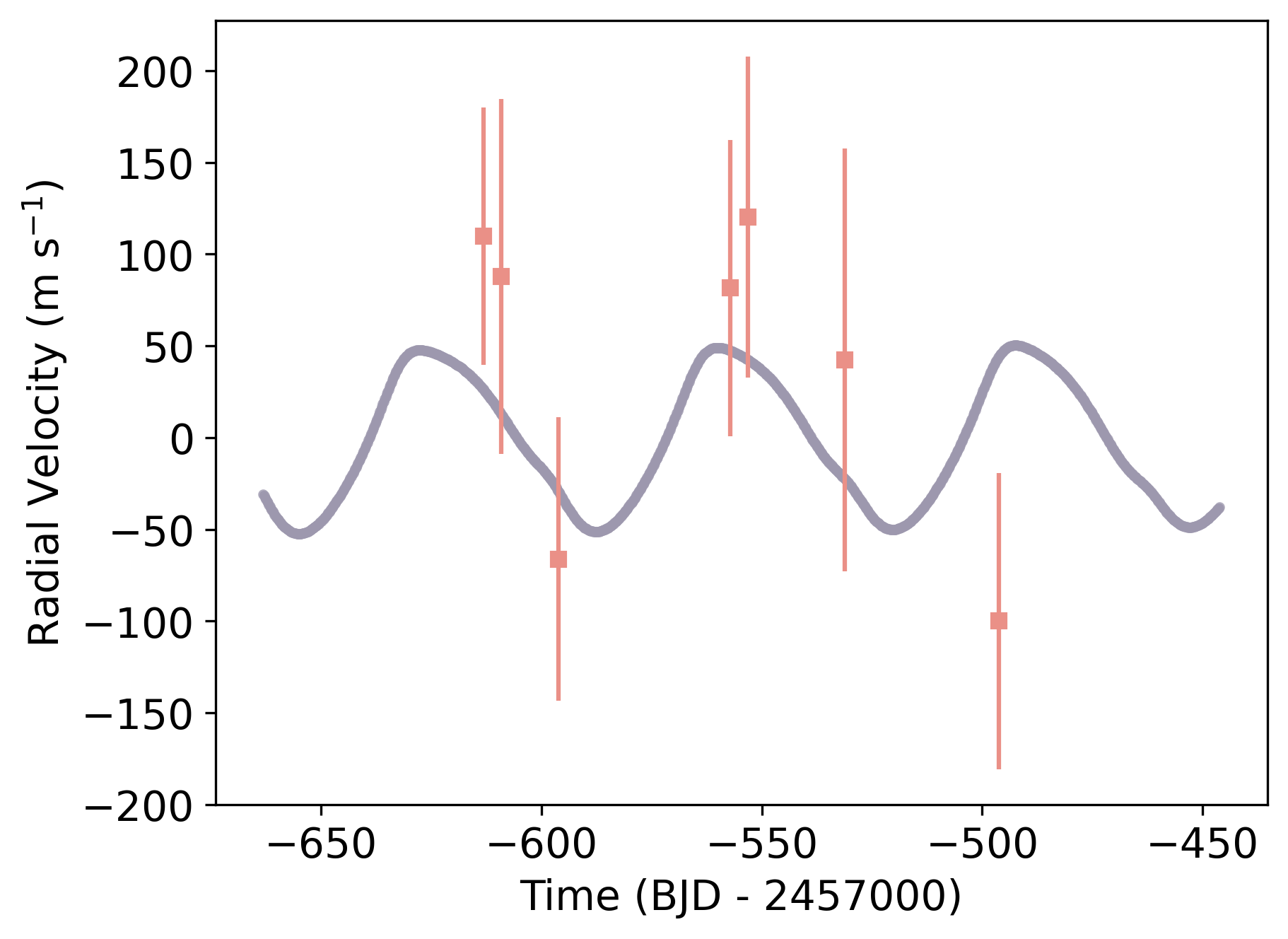}\hfill

\caption{\textbf{Radial velocity observations of KOI-134 with TRES.} TRES radial velocity measurements of KOI-134 taken during the early 2013 observing season, overplotted with a {\tt radvel} \cite{radvel} model propagated for 50 days forwards and backwards in time. Each point shows the median RV value with 1-$\sigma$ errorbars. We omit the two RV observations from the previous observing season, as these contain outliers, and considering the seasonal variation of the instrument is beyond the scope of this work.}
\label{fig:RVcurve}

\end{figure}

\begin{table}[h]
\caption{
 Transit Times of  KOI-134\,b from \textit{Kepler}
  \label{tab:transittimes}
}
\begin{tabular}{cccccccccc}
\toprule
Epoch & Time (BJD-2457000) & $\sigma_{epoch}$ & Duration (Hr) & $\sigma_{dur}$ (Hr) \\
\midrule
1 & -2013.7511 & 1.1e-03 & 11.449 & 0.096\\
2 & -1946.63556 & 7.2e-04 & 11.580 & 0.078\\
3 & -1879.45279 & 7.3e-04 & 11.377 & 0.057\\
4 & -1812.19897 & 7.9e-04 & 11.595 & 0.074\\
5 & -1744.86761 & 9.3e-04 & 11.644 & 0.072\\
6 & -1677.452204 & 7.09e-04 & 11.530 & 0.064\\
7 & -1609.96144 & 7.6e-04 & 11.569 & 0.058\\
8 & -1542.3899340 & 7.03e-04 & 11.656 & 0.065\\
9 & -1474.74328 & 7.5e-04 & 11.572 & 0.073\\
10 & -1407.036146 & 9.04e-04 & 11.650 & 0.080\\
11 & -1339.29612 & 8.7e-04 & 11.638 & 0.061\\
12 & -1271.54801 & 8.8e-04 & 11.599 & 0.061\\
13 & -1203.82174 & 7.9e-04 & 11.595 & 0.085\\
14 & -1136.146997 & 7.02e-04 & 11.614 & 0.080\\
16 & -1001.00193 & 9.6e-04 & 11.504 & 0.060\\
17 & -933.54480 & 8.8e-04 & 11.491 & 0.075\\
18 & -866.1671 & 1.4e-03 & 11.37 & 0.11\\
19 & -798.86548 & 8.6e-04 & 11.360 & 0.087\\
21 & -664.48537 & 9.3e-04 & 11.412 & 0.077\\
22 & -597.39243 & 8.7e-04 & 11.367 & 0.082\\
\botrule
\end{tabular}
\end{table}

\begin{figure*}[htp]
\centering
\includegraphics[width=1\textwidth]{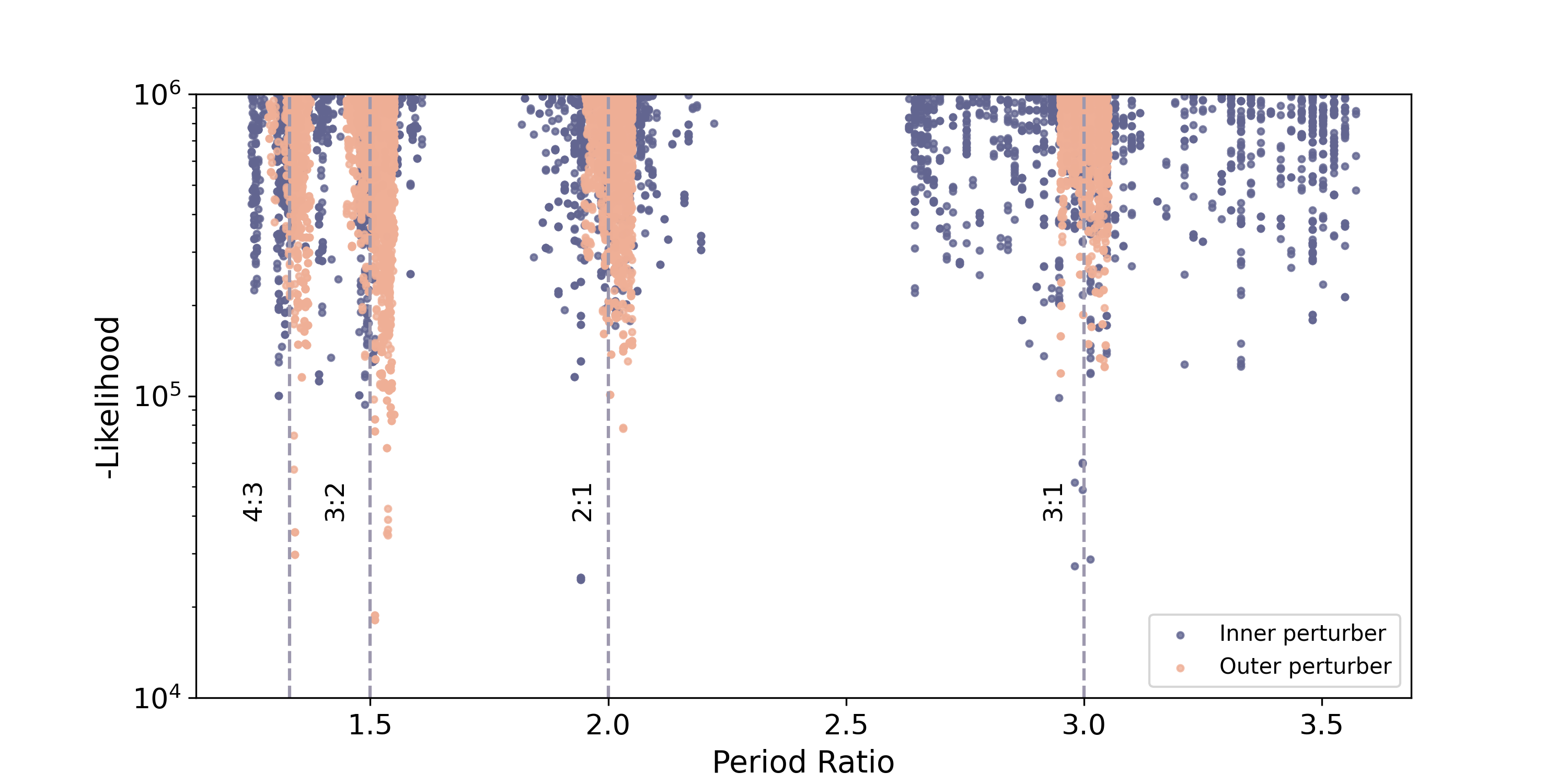}\hfill

\caption{\textbf{Grid search results depicting likelihood as a function of orbital resonance.} {Likelihood as a function of period ratio for each of the 8 orbital resonances explored during our grid search. The purple points denote an inner perturber, where KOI-134\,c is interior to KOI-134\,b, and the displayed period ratio is then {$P_{b}/P_{c}$}. The coral-colored points indicate the opposite case. We plot the negative likelihood in order to scale the y-axis logarithmically, therefore the lowermost points represent the highest likelihood. The values that generate the highest likelihoods are then used as starting positions for a more thorough MCMC fitting routine.}}
\label{fig:pratio}

\end{figure*}

\begin{figure}[htp]
\centering
\includegraphics[width=0.8\linewidth]{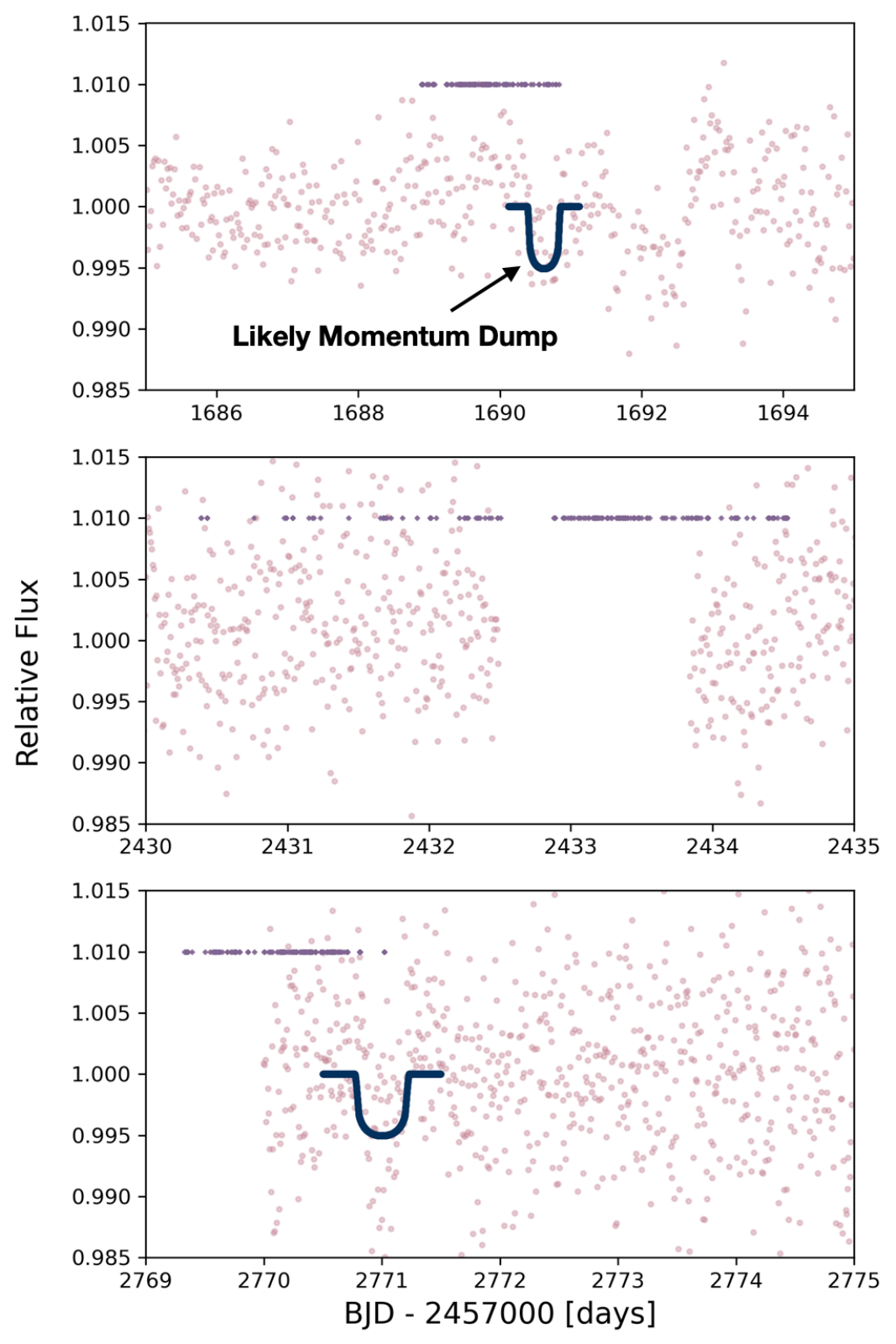}

\caption{\textbf{Predicted transits of KOI-134\,b compared to TESS data.} Predicted transit centers of KOI-134\,b (purple dots) overlaid with segments of \textit{TESS} light curves. The transit models in dark blue are created with best-fit planet parameters of KOI-134\,b, denoting the candidate transit events found in the \textit{TESS} light curves. The purple points appear to span different lengths because the x-axes do not share the same scale.}
\label{fig:tess}

\end{figure}

\clearpage
\newpage

\renewcommand\refname{Supplementary References}



\begin{thebibliography}{10}
\expandafter\ifx\csname url\endcsname\relax
  \def\url#1{\burl{#1}}\fi
\expandafter\ifx\csname urlprefix\endcsname\relax\def\urlprefix{URL }\fi
\providecommand{\bibinfo}[2]{#2}
\providecommand{\eprint}[2][]{\url{#2}}
\providecommand{\doi}[1]{\url{https://doi.org/#1}}
\bibcommenthead

\bibitem{Rivera:2010}
\bibinfo{author}{{Rivera}, E.~J.} \emph{et~al.}
\newblock \bibinfo{title}{{The Lick-Carnegie Exoplanet Survey: a Uranus-Mass Fourth Planet for GJ 876 in an Extrasolar Laplace Configuration}}.
\newblock \emph{\bibinfo{journal}{\apj}} \textbf{\bibinfo{volume}{719}}, \bibinfo{pages}{890--899} (\bibinfo{year}{2010}).

\bibitem{Sanchis-Ojeda:2012}
\bibinfo{author}{{Sanchis-Ojeda}, R.} \emph{et~al.}
\newblock \bibinfo{title}{{Alignment of the stellar spin with the orbits of a three-planet system}}.
\newblock \emph{\bibinfo{journal}{\nat}} \textbf{\bibinfo{volume}{487}}, \bibinfo{pages}{449--453} (\bibinfo{year}{2012}).

\bibitem{Nesvorny:2012}
\bibinfo{author}{{Nesvorn{\'y}}, D.} \emph{et~al.}
\newblock \bibinfo{title}{{The Detection and Characterization of a Nontransiting Planet by Transit Timing Variations}}.
\newblock \emph{\bibinfo{journal}{Science}} \textbf{\bibinfo{volume}{336}}, \bibinfo{pages}{1133} (\bibinfo{year}{2012}).

\bibitem{Huber:2013}
\bibinfo{author}{{Huber}, D.} \emph{et~al.}
\newblock \bibinfo{title}{{Stellar Spin-Orbit Misalignment in a Multiplanet System}}.
\newblock \emph{\bibinfo{journal}{Science}} \textbf{\bibinfo{volume}{342}}, \bibinfo{pages}{331--334} (\bibinfo{year}{2013}).

\bibitem{Almenara:2015}
\bibinfo{author}{{Almenara}, J.~M.} \emph{et~al.}
\newblock \bibinfo{title}{{Absolute masses and radii determination in multiplanetary systems without stellar models}}.
\newblock \emph{\bibinfo{journal}{\mnras}} \textbf{\bibinfo{volume}{453}}, \bibinfo{pages}{2644--2652} (\bibinfo{year}{2015}).

\bibitem{Hamann:2019}
\bibinfo{author}{{Hamann}, A.}, \bibinfo{author}{{Montet}, B.~T.}, \bibinfo{author}{{Fabrycky}, D.~C.}, \bibinfo{author}{{Agol}, E.} \& \bibinfo{author}{{Kruse}, E.}
\newblock \bibinfo{title}{{K2-146: Discovery of Planet c, Precise Masses from Transit Timing, and Observed Precession}}.
\newblock \emph{\bibinfo{journal}{\aj}} \textbf{\bibinfo{volume}{158}}, \bibinfo{pages}{133} (\bibinfo{year}{2019}).

\bibitem{Johnson:2022}
\bibinfo{author}{{Johnson}, M.~C.} \emph{et~al.}
\newblock \bibinfo{title}{{An Aligned Orbit for the Young Planet V1298 Tau b}}.
\newblock \emph{\bibinfo{journal}{\aj}} \textbf{\bibinfo{volume}{163}}, \bibinfo{pages}{247} (\bibinfo{year}{2022}).

\bibitem{Feinstein:2021}
\bibinfo{author}{{Feinstein}, A.~D.} \emph{et~al.}
\newblock \bibinfo{title}{{H-alpha and Ca II Infrared Triplet Variations During a Transit of the 23 Myr Planet V1298 Tau c}}.
\newblock \emph{\bibinfo{journal}{\aj}} \textbf{\bibinfo{volume}{162}}, \bibinfo{pages}{213} (\bibinfo{year}{2021}).

\bibitem{GoldreichTremaine:1980}
\bibinfo{author}{{Goldreich}, P.} \& \bibinfo{author}{{Tremaine}, S.}
\newblock \bibinfo{title}{{Disk-satellite interactions.}}
\newblock \emph{\bibinfo{journal}{\apj}} \textbf{\bibinfo{volume}{241}}, \bibinfo{pages}{425--441} (\bibinfo{year}{1980}).

\bibitem{Lissauer:2011}
\bibinfo{author}{{Lissauer}, J.~J.} \emph{et~al.}
\newblock \bibinfo{title}{{Architecture and Dynamics of Kepler's Candidate Multiple Transiting Planet Systems}}.
\newblock \emph{\bibinfo{journal}{\apjs}} \textbf{\bibinfo{volume}{197}}, \bibinfo{pages}{8} (\bibinfo{year}{2011}).

\bibitem{Steffen:2010}
\bibinfo{author}{{Steffen}, J.~H.} \emph{et~al.}
\newblock \bibinfo{title}{{Five Kepler Target Stars That Show Multiple Transiting Exoplanet Candidates}}.
\newblock \emph{\bibinfo{journal}{\apj}} \textbf{\bibinfo{volume}{725}}, \bibinfo{pages}{1226--1241} (\bibinfo{year}{2010}).

\bibitem{Fang:2013}
\bibinfo{author}{{Fang}, J.} \& \bibinfo{author}{{Margot}, J.-L.}
\newblock \bibinfo{title}{{Are Planetary Systems Filled to Capacity? A Study Based on Kepler Results}}.
\newblock \emph{\bibinfo{journal}{\apj}} \textbf{\bibinfo{volume}{767}}, \bibinfo{pages}{115} (\bibinfo{year}{2013}).

\bibitem{Fabrycky:2014}
\bibinfo{author}{{Fabrycky}, D.~C.} \emph{et~al.}
\newblock \bibinfo{title}{{Architecture of Kepler's Multi-transiting Systems. II. New Investigations with Twice as Many Candidates}}.
\newblock \emph{\bibinfo{journal}{\apj}} \textbf{\bibinfo{volume}{790}}, \bibinfo{pages}{146} (\bibinfo{year}{2014}).

\bibitem{Marzari:2002}
\bibinfo{author}{{Marzari}, F.} \& \bibinfo{author}{{Weidenschilling}, S.~J.}
\newblock \bibinfo{title}{{Eccentric Extrasolar Planets: The Jumping Jupiter Model}}.
\newblock \emph{\bibinfo{journal}{\icarus}} \textbf{\bibinfo{volume}{156}}, \bibinfo{pages}{570--579} (\bibinfo{year}{2002}).

\bibitem{Chatterjee:2008}
\bibinfo{author}{{Chatterjee}, S.}, \bibinfo{author}{{Ford}, E.~B.}, \bibinfo{author}{{Matsumura}, S.} \& \bibinfo{author}{{Rasio}, F.~A.}
\newblock \bibinfo{title}{{Dynamical Outcomes of Planet-Planet Scattering}}.
\newblock \emph{\bibinfo{journal}{\apj}} \textbf{\bibinfo{volume}{686}}, \bibinfo{pages}{580--602} (\bibinfo{year}{2008}).

\bibitem{JuricTremaine:2008}
\bibinfo{author}{{Juri{\'c}}, M.} \& \bibinfo{author}{{Tremaine}, S.}
\newblock \bibinfo{title}{{Dynamical Origin of Extrasolar Planet Eccentricity Distribution}}.
\newblock \emph{\bibinfo{journal}{\apj}} \textbf{\bibinfo{volume}{686}}, \bibinfo{pages}{603--620} (\bibinfo{year}{2008}).

\bibitem{DawsonChiang2014}
\bibinfo{author}{{Dawson}, R.~I.} \& \bibinfo{author}{{Chiang}, E.}
\newblock \bibinfo{title}{{A class of warm Jupiters with mutually inclined, apsidally misaligned close friends}}.
\newblock \emph{\bibinfo{journal}{Science}} \textbf{\bibinfo{volume}{346}}, \bibinfo{pages}{212--216} (\bibinfo{year}{2014}).

\bibitem{McArthur:2010}
\bibinfo{author}{{McArthur}, B.~E.} \emph{et~al.}
\newblock \bibinfo{title}{{New Observational Constraints on the {\ensuremath{\upsilon}} Andromedae System with Data from the Hubble Space Telescope and Hobby-Eberly Telescope}}.
\newblock \emph{\bibinfo{journal}{\apj}} \textbf{\bibinfo{volume}{715}}, \bibinfo{pages}{1203--1220} (\bibinfo{year}{2010}).

\bibitem{Dawson:2014}
\bibinfo{author}{{Dawson}, R.~I.} \emph{et~al.}
\newblock \bibinfo{title}{{Large Eccentricity, Low Mutual Inclination: The Three-dimensional Architecture of a Hierarchical System of Giant Planets}}.
\newblock \emph{\bibinfo{journal}{\apj}} \textbf{\bibinfo{volume}{791}}, \bibinfo{pages}{89} (\bibinfo{year}{2014}).

\bibitem{MillsFabrycky:2017}
\bibinfo{author}{{Mills}, S.~M.} \& \bibinfo{author}{{Fabrycky}, D.~C.}
\newblock \bibinfo{title}{{Kepler-108: A Mutually Inclined Giant Planet System}}.
\newblock \emph{\bibinfo{journal}{\aj}} \textbf{\bibinfo{volume}{153}}, \bibinfo{pages}{45} (\bibinfo{year}{2017}).

\bibitem{rebound}
\bibinfo{author}{{Rein}, H.} \& \bibinfo{author}{{Liu}, S.-F.}
\newblock \bibinfo{title}{{REBOUND: an open-source multi-purpose N-body code for collisional dynamics}}.
\newblock \emph{\bibinfo{journal}{\aap}} \textbf{\bibinfo{volume}{537}}, \bibinfo{pages}{A128} (\bibinfo{year}{2012}).

\bibitem{Korth:2023}
\bibinfo{author}{{Korth}, J.} \emph{et~al.}
\newblock \bibinfo{title}{{TOI-1130: A photodynamical analysis of a hot Jupiter in resonance with an inner low-mass planet}}.
\newblock \emph{\bibinfo{journal}{\aap}} \textbf{\bibinfo{volume}{675}}, \bibinfo{pages}{A115} (\bibinfo{year}{2023}).

\bibitem{Korth:2024}
\bibinfo{author}{{Korth}, J.} \emph{et~al.}
\newblock \bibinfo{title}{{TOI-1408: Discovery and Photodynamical Modeling of a Small Inner Companion to a Hot Jupiter Revealed by Transit Timing Variations}}.
\newblock \emph{\bibinfo{journal}{\apjl}} \textbf{\bibinfo{volume}{971}}, \bibinfo{pages}{L28} (\bibinfo{year}{2024}).

\bibitem{ricker}
\bibinfo{author}{{Ricker}, G.~R.} \emph{et~al.}
\newblock \bibinfo{title}{{Transiting Exoplanet Survey Satellite (TESS)}}.
\newblock \emph{\bibinfo{journal}{\jatis}} \textbf{\bibinfo{volume}{1}}, \bibinfo{pages}{014003} (\bibinfo{year}{2015}).

\bibitem{Dawson:2019}
\bibinfo{author}{{Dawson}, R.~I.} \emph{et~al.}
\newblock \bibinfo{title}{{TOI-216b and TOI-216 c: Two Warm, Large Exoplanets in or Slightly Wide of the 2:1 Orbital Resonance}}.
\newblock \emph{\bibinfo{journal}{\aj}} \textbf{\bibinfo{volume}{158}}, \bibinfo{pages}{65} (\bibinfo{year}{2019}).

\bibitem{Dawson:2021}
\bibinfo{author}{{Dawson}, R.~I.} \emph{et~al.}
\newblock \bibinfo{title}{{Precise Transit and Radial-velocity Characterization of a Resonant Pair: The Warm Jupiter TOI-216c and Eccentric Warm Neptune TOI-216b}}.
\newblock \emph{\bibinfo{journal}{\aj}} \textbf{\bibinfo{volume}{161}}, \bibinfo{pages}{161} (\bibinfo{year}{2021}).

\bibitem{Nesvorny:2022}
\bibinfo{author}{{Nesvorn{\'y}}, D.}, \bibinfo{author}{{Chrenko}, O.} \& \bibinfo{author}{{Flock}, M.}
\newblock \bibinfo{title}{{TOI-216: Resonant Constraints on Planet Migration}}.
\newblock \emph{\bibinfo{journal}{\apj}} \textbf{\bibinfo{volume}{925}}, \bibinfo{pages}{38} (\bibinfo{year}{2022}).

\bibitem{Nesvorny:2013}
\bibinfo{author}{{Nesvorn{\'y}}, D.} \emph{et~al.}
\newblock \bibinfo{title}{{KOI-142, The King of Transit Variations, is a Pair of Planets near the 2:1 Resonance}}.
\newblock \emph{\bibinfo{journal}{\apj}} \textbf{\bibinfo{volume}{777}}, \bibinfo{pages}{3} (\bibinfo{year}{2013}).

\bibitem{Lithwick2012}
\bibinfo{author}{{Lithwick}, Y.}, \bibinfo{author}{{Xie}, J.} \& \bibinfo{author}{{Wu}, Y.}
\newblock \bibinfo{title}{{Extracting Planet Mass and Eccentricity from TTV Data}}.
\newblock \emph{\bibinfo{journal}{\apj}} \textbf{\bibinfo{volume}{761}}, \bibinfo{pages}{122} (\bibinfo{year}{2012}).

\bibitem{Deck:2013}
\bibinfo{author}{{Deck}, K.~M.}, \bibinfo{author}{{Payne}, M.} \& \bibinfo{author}{{Holman}, M.~J.}
\newblock \bibinfo{title}{{First-order Resonance Overlap and the Stability of Close Two-planet Systems}}.
\newblock \emph{\bibinfo{journal}{\apj}} \textbf{\bibinfo{volume}{774}}, \bibinfo{pages}{129} (\bibinfo{year}{2013}).

\bibitem{Murray:1999}
\bibinfo{author}{{Murray}, C.~D.} \& \bibinfo{author}{{Dermott}, S.~F.}
\newblock \emph{\bibinfo{title}{{Solar System Dynamics}}}, \bibinfo{publisher}{{Cambridge University Press}}  (\bibinfo{year}{1999}).

\bibitem{ThommesLissauer:2003}
\bibinfo{author}{{Thommes}, E.~W.} \& \bibinfo{author}{{Lissauer}, J.~J.}
\newblock \bibinfo{title}{{Resonant Inclination Excitation of Migrating Giant Planets}}.
\newblock \emph{\bibinfo{journal}{\apj}} \textbf{\bibinfo{volume}{597}}, \bibinfo{pages}{566--580} (\bibinfo{year}{2003}).

\bibitem{Cresswell:2008}
\bibinfo{author}{{Cresswell}, P.} \& \bibinfo{author}{{Nelson}, R.~P.}
\newblock \bibinfo{title}{{Three-dimensional simulations of multiple protoplanets embedded in a protostellar disc}}.
\newblock \emph{\bibinfo{journal}{\aap}} \textbf{\bibinfo{volume}{482}}, \bibinfo{pages}{677--690} (\bibinfo{year}{2008}).

\bibitem{jenkins2010}
\bibinfo{author}{{Jenkins}, J.~M.} \emph{et~al.}
\newblock \bibinfo{title}{{Overview of the Kepler Science Processing Pipeline}}.
\newblock \emph{\bibinfo{journal}{\apjl}} \textbf{\bibinfo{volume}{713}}, \bibinfo{pages}{L87--L91} (\bibinfo{year}{2010}).

\bibitem{Coughlin:2016}
\bibinfo{author}{{Coughlin}, J.~L.} \emph{et~al.}
\newblock \bibinfo{title}{{Planetary Candidates Observed by Kepler. VII. The First Fully Uniform Catalog Based on the Entire 48-month Data Set (Q1-Q17 DR24)}}.
\newblock \emph{\bibinfo{journal}{\apjs}} \textbf{\bibinfo{volume}{224}}, \bibinfo{pages}{12} (\bibinfo{year}{2016}).

\bibitem{Borucki:2011}
\bibinfo{author}{{Borucki}, W.~J.} \emph{et~al.}
\newblock \bibinfo{title}{{Characteristics of Planetary Candidates Observed by Kepler. II. Analysis of the First Four Months of Data}}.
\newblock \emph{\bibinfo{journal}{\apj}} \textbf{\bibinfo{volume}{736}}, \bibinfo{pages}{19} (\bibinfo{year}{2011}).

\bibitem{Thompson:2018}
\bibinfo{author}{{Thompson}, S.~E.} \emph{et~al.}
\newblock \bibinfo{title}{{Planetary Candidates Observed by Kepler. VIII. A Fully Automated Catalog with Measured Completeness and Reliability Based on Data Release 25}}.
\newblock \emph{\bibinfo{journal}{\apjs}} \textbf{\bibinfo{volume}{235}}, \bibinfo{pages}{38} (\bibinfo{year}{2018}).

\bibitem{Bryson:2017}
\bibinfo{author}{{Bryson}, S.~T.} \emph{et~al.}
\newblock \bibinfo{title}{{The Kepler Certified False Positive Table}}.
\newblock \bibinfo{howpublished}{Kepler Science Document KSCI-19093-003, id. 12. Edited by Michael R. Haas and Natalie M. Batalha} (\bibinfo{year}{2017}).

\bibitem{Vanderburg:2020}
\bibinfo{author}{{Vanderburg}, A.} \emph{et~al.}
\newblock \bibinfo{title}{{A Habitable-zone Earth-sized Planet Rescued from False Positive Status}}.
\newblock \emph{\bibinfo{journal}{\apjl}} \textbf{\bibinfo{volume}{893}}, \bibinfo{pages}{L27} (\bibinfo{year}{2020}).

\bibitem{TRES}
\bibinfo{author}{F{\H{u}}r{\'e}sz, G.}, \bibinfo{author}{Szentgyorgyi, A.~H.} \& \bibinfo{author}{Meibom, S.}
\newblock \bibinfo{title}{Precision of radial velocity surveys using multiobject spectrographs --- experiences with hectochelle} \bibinfo{pages}{287--289} (\bibinfo{year}{2008}).

\bibitem{ZhouLSD}
\bibinfo{author}{{Zhou}, G.} \emph{et~al.}
\newblock \bibinfo{title}{{A Well-aligned Orbit for the 45 Myr-old Transiting Neptune DS Tuc Ab}}.
\newblock \emph{\bibinfo{journal}{\apjl}} \textbf{\bibinfo{volume}{892}}, \bibinfo{pages}{L21} (\bibinfo{year}{2020}).

\bibitem{HIRES}
\bibinfo{author}{{Vogt}, S.~S.} \emph{et~al.}
\newblock \bibinfo{editor}{{Crawford}, D.~L.} \& \bibinfo{editor}{{Craine}, E.~R.} (eds) \emph{\bibinfo{title}{{HIRES: the high-resolution echelle spectrometer on the Keck 10-m Telescope}}}.
\newblock (eds \bibinfo{editor}{{Crawford}, D.~L.} \& \bibinfo{editor}{{Craine}, E.~R.}) \emph{\bibinfo{booktitle}{Instrumentation in Astronomy VIII}}, Vol. \bibinfo{volume}{2198} of \emph{\bibinfo{series}{Society of Photo-Optical Instrumentation Engineers (SPIE) Conference Series}}, \bibinfo{pages}{362} (\bibinfo{year}{1994}).

\bibitem{Tull:1994}
\bibinfo{author}{{Tull}, R.~G.}, \bibinfo{author}{{MacQueen}, P.}, \bibinfo{author}{{Sneden}, C.} \& \bibinfo{author}{{Lambert}, D.~L.}
\newblock \bibinfo{editor}{{Pyper}, D.~M.} \& \bibinfo{editor}{{Angione}, R.~J.} (eds) \emph{\bibinfo{title}{{The McDonald 2.7-in Echelle Spectrometer}}}.
\newblock (eds \bibinfo{editor}{{Pyper}, D.~M.} \& \bibinfo{editor}{{Angione}, R.~J.}) \emph{\bibinfo{booktitle}{Optical Astronomy from the Earth and Moon}}, Vol.~\bibinfo{volume}{55} of \emph{\bibinfo{series}{Astronomical Society of the Pacific Conference Series}}, \bibinfo{pages}{148} (\bibinfo{year}{1994}).

\bibitem{Buchhave:2012}
\bibinfo{author}{{Buchhave}, L.~A.} \emph{et~al.}
\newblock \bibinfo{title}{{An abundance of small exoplanets around stars with a wide range of metallicities}}.
\newblock \emph{\bibinfo{journal}{\nat}} \textbf{\bibinfo{volume}{486}}, \bibinfo{pages}{375--377} (\bibinfo{year}{2012}).

\bibitem{Valenti:1996}
\bibinfo{author}{{Valenti}, J.~A.} \& \bibinfo{author}{{Piskunov}, N.}
\newblock \bibinfo{title}{{Spectroscopy made easy: A new tool for fitting observations with synthetic spectra.}}
\newblock \emph{\bibinfo{journal}{\aaps}} \textbf{\bibinfo{volume}{118}}, \bibinfo{pages}{595--603} (\bibinfo{year}{1996}).

\bibitem{Baranec:2013}
\bibinfo{author}{{Baranec}, C.} \emph{et~al.}
\newblock \bibinfo{title}{{Bringing the Visible Universe into Focus with Robo-AO}}.
\newblock \emph{\bibinfo{journal}{\jve}} \textbf{\bibinfo{volume}{72}}, \bibinfo{pages}{50021} (\bibinfo{year}{2013}).

\bibitem{Baranec:2014}
\bibinfo{author}{{Baranec}, C.} \emph{et~al.}
\newblock \bibinfo{title}{{High-efficiency Autonomous Laser Adaptive Optics}}.
\newblock \emph{\bibinfo{journal}{\apjl}} \textbf{\bibinfo{volume}{790}}, \bibinfo{pages}{L8} (\bibinfo{year}{2014}).

\bibitem{Ziegler:2017}
\bibinfo{author}{{Ziegler}, C.} \emph{et~al.}
\newblock \bibinfo{title}{{Robo-AO Kepler Planetary Candidate Survey. III. Adaptive Optics Imaging of 1629 Kepler Exoplanet Candidate Host Stars}}.
\newblock \emph{\bibinfo{journal}{\aj}} \textbf{\bibinfo{volume}{153}}, \bibinfo{pages}{66} (\bibinfo{year}{2017}).

\bibitem{Tayar:2022}
\bibinfo{author}{{Tayar}, J.}, \bibinfo{author}{{Claytor}, Z.~R.}, \bibinfo{author}{{Huber}, D.} \& \bibinfo{author}{{van Saders}, J.}
\newblock \bibinfo{title}{{A Guide to Realistic Uncertainties on the Fundamental Properties of Solar-type Exoplanet Host Stars}}.
\newblock \emph{\bibinfo{journal}{\apj}} \textbf{\bibinfo{volume}{927}}, \bibinfo{pages}{31} (\bibinfo{year}{2022}).

\bibitem{Eastman:2019}
\bibinfo{author}{{Eastman}, J.~D.} \emph{et~al.}
\newblock \bibinfo{title}{{EXOFASTv2: A public, generalized, publication-quality exoplanet modeling code}}.
\newblock \emph{\bibinfo{journal}{arXiv e-prints}} \bibinfo{pages}{arXiv:1907.09480} (\bibinfo{year}{2019}).

\bibitem{radvel}
\bibinfo{author}{{Fulton}, B.~J.}, \bibinfo{author}{{Petigura}, E.~A.}, \bibinfo{author}{{Blunt}, S.} \& \bibinfo{author}{{Sinukoff}, E.}
\newblock \bibinfo{title}{{RadVel: The Radial Velocity Modeling Toolkit}}.
\newblock \emph{\bibinfo{journal}{\pasp}} \textbf{\bibinfo{volume}{130}}, \bibinfo{pages}{044504} (\bibinfo{year}{2018}).

\bibitem{Foreman-Mackey2013}
\bibinfo{author}{{Foreman-Mackey}, D.}, \bibinfo{author}{{Hogg}, D.~W.}, \bibinfo{author}{{Lang}, D.} \& \bibinfo{author}{{Goodman}, J.}
\newblock \bibinfo{title}{{emcee: The MCMC Hammer}}.
\newblock \emph{\bibinfo{journal}{\pasp}} \textbf{\bibinfo{volume}{125}}, \bibinfo{pages}{306} (\bibinfo{year}{2013}).

\bibitem{Kreidberg(2015)}
\bibinfo{author}{{Kreidberg}, L.}
\newblock \bibinfo{title}{{batman: BAsic Transit Model cAlculatioN in Python}}.
\newblock \emph{\bibinfo{journal}{\pasp}} \textbf{\bibinfo{volume}{127}}, \bibinfo{pages}{1161} (\bibinfo{year}{2015}).

\bibitem{Murphy:2012}
\bibinfo{author}{{Murphy}, S.~J.}
\newblock \bibinfo{title}{{An examination of some characteristics of Kepler short- and long-cadence data}}.
\newblock \emph{\bibinfo{journal}{\mnras}} \textbf{\bibinfo{volume}{422}}, \bibinfo{pages}{665--671} (\bibinfo{year}{2012}).

\bibitem{MISTModels}
\bibinfo{author}{{Choi}, J.} \emph{et~al.}
\newblock \bibinfo{title}{{Mesa Isochrones and Stellar Tracks (MIST). I. Solar-scaled Models}}.
\newblock \emph{\bibinfo{journal}{\apj}} \textbf{\bibinfo{volume}{823}}, \bibinfo{pages}{102} (\bibinfo{year}{2016}).

\bibitem{Green:2018}
\bibinfo{author}{{Green}, G.~M.}
\newblock \bibinfo{title}{{dustmaps: A Python interface for maps of interstellar dust}}.
\newblock \emph{\bibinfo{journal}{\joss}} \textbf{\bibinfo{volume}{3}}, \bibinfo{pages}{695} (\bibinfo{year}{2018}).

\bibitem{Schlegel:1998}
\bibinfo{author}{{Schlegel}, D.~J.}, \bibinfo{author}{{Finkbeiner}, D.~P.} \& \bibinfo{author}{{Davis}, M.}
\newblock \bibinfo{title}{{Maps of Dust Infrared Emission for Use in Estimation of Reddening and Cosmic Microwave Background Radiation Foregrounds}}.
\newblock \emph{\bibinfo{journal}{\apj}} \textbf{\bibinfo{volume}{500}}, \bibinfo{pages}{525--553} (\bibinfo{year}{1998}).

\bibitem{Schlafly:2011}
\bibinfo{author}{{Schlafly}, E.~F.} \& \bibinfo{author}{{Finkbeiner}, D.~P.}
\newblock \bibinfo{title}{{Measuring Reddening with Sloan Digital Sky Survey Stellar Spectra and Recalibrating SFD}}.
\newblock \emph{\bibinfo{journal}{\apj}} \textbf{\bibinfo{volume}{737}}, \bibinfo{pages}{103} (\bibinfo{year}{2011}).

\bibitem{Kipping:2013}
\bibinfo{author}{{Kipping}, D.~M.}
\newblock \bibinfo{title}{{Efficient, uninformative sampling of limb darkening coefficients for two-parameter laws}}.
\newblock \emph{\bibinfo{journal}{\mnras}} \textbf{\bibinfo{volume}{435}}, \bibinfo{pages}{2152--2160} (\bibinfo{year}{2013}).

\bibitem{SeagerMallenOrnelas2003}
\bibinfo{author}{{Seager}, S.} \& \bibinfo{author}{{Mall{\'e}n-Ornelas}, G.}
\newblock \bibinfo{title}{{A Unique Solution of Planet and Star Parameters from an Extrasolar Planet Transit Light Curve}}.
\newblock \emph{\bibinfo{journal}{\apj}} \textbf{\bibinfo{volume}{585}}, \bibinfo{pages}{1038--1055} (\bibinfo{year}{2003}).

\bibitem{ias15}
\bibinfo{author}{{Rein}, H.} \& \bibinfo{author}{{Spiegel}, D.~S.}
\newblock \bibinfo{title}{{IAS15: a fast, adaptive, high-order integrator for gravitational dynamics, accurate to machine precision over a billion orbits}}.
\newblock \emph{\bibinfo{journal}{\mnras}} \textbf{\bibinfo{volume}{446}}, \bibinfo{pages}{1424--1437} (\bibinfo{year}{2015}).

\bibitem{Kipping:2010}
\bibinfo{author}{{Kipping}, D.~M.}
\newblock \bibinfo{title}{{Investigations of approximate expressions for the transit duration}}.
\newblock \emph{\bibinfo{journal}{\mnras}} \textbf{\bibinfo{volume}{407}}, \bibinfo{pages}{301--313} (\bibinfo{year}{2010}).

\bibitem{Korth:2020}
\bibinfo{author}{Korth, J.}
\newblock \emph{\bibinfo{title}{{Characterization of extrasolar multi-planet systems by transit timing variation}}}.
\newblock Ph.D. thesis, \bibinfo{school}{Universit{\"a}t zu K{\"o}ln} (\bibinfo{year}{2020}).
\newblock \urlprefix\url{https://kups.ub.uni-koeln.de/11289/}.

\bibitem{NesvornyVokrouhlicky:2016}
\bibinfo{author}{{Nesvorn{\'y}}, D.} \& \bibinfo{author}{{Vokrouhlick{\'y}}, D.}
\newblock \bibinfo{title}{{Dynamics and Transit Variations of Resonant Exoplanets}}.
\newblock \emph{\bibinfo{journal}{\apj}} \textbf{\bibinfo{volume}{823}}, \bibinfo{pages}{72} (\bibinfo{year}{2016}).

\bibitem{LithwickWu:2012}
\bibinfo{author}{{Lithwick}, Y.} \& \bibinfo{author}{{Wu}, Y.}
\newblock \bibinfo{title}{{Resonant Repulsion of Kepler Planet Pairs}}.
\newblock \emph{\bibinfo{journal}{\apjl}} \textbf{\bibinfo{volume}{756}}, \bibinfo{pages}{L11} (\bibinfo{year}{2012}).

\bibitem{Aarseth:2003}
\bibinfo{author}{{Aarseth}, S.~J.}
\newblock \emph{\bibinfo{title}{{Gravitational N-Body Simulations}}}  (\bibinfo{year}{2003}).

\bibitem{Faridani23}
\bibinfo{author}{{Faridani}, T.~H.}, \bibinfo{author}{{Naoz}, S.}, \bibinfo{author}{{Li}, G.} \& \bibinfo{author}{{Inzunza}, N.}
\newblock \bibinfo{title}{{Let's Sweep: The Effect of Evolving J $_{2}$ on the Resonant Structure of a Three-planet System}}.
\newblock \emph{\bibinfo{journal}{\apj}} \textbf{\bibinfo{volume}{956}}, \bibinfo{pages}{90} (\bibinfo{year}{2023}).

\bibitem{Faridani24}
\bibinfo{author}{{Faridani}, T.}, \bibinfo{author}{{Naoz}, S.}, \bibinfo{author}{{Li}, G.}, \bibinfo{author}{{Rice}, M.} \& \bibinfo{author}{{Inzunza}, N.}
\newblock \bibinfo{title}{{More Likely Than You Think: Inclination-Driving Secular Resonances are Common in Known Exoplanet Systems}}.
\newblock \emph{\bibinfo{journal}{arXiv e-prints}} \bibinfo{pages}{arXiv:2406.09359} (\bibinfo{year}{2024}).

\bibitem{Nagasawa03}
\bibinfo{author}{{Nagasawa}, M.}, \bibinfo{author}{{Lin}, D.~N.~C.} \& \bibinfo{author}{{Ida}, S.}
\newblock \bibinfo{title}{{Eccentricity Evolution of Extrasolar Multiple Planetary Systems Due to the Depletion of Nascent Protostellar Disks}}.
\newblock \emph{\bibinfo{journal}{\apj}} \textbf{\bibinfo{volume}{586}}, \bibinfo{pages}{1374--1393} (\bibinfo{year}{2003}).

\bibitem{Petrovich20}
\bibinfo{author}{{Petrovich}, C.}, \bibinfo{author}{{Mu{\~n}oz}, D.~J.}, \bibinfo{author}{{Kratter}, K.~M.} \& \bibinfo{author}{{Malhotra}, R.}
\newblock \bibinfo{title}{{A Disk-driven Resonance as the Origin of High Inclinations of Close-in Planets}}.
\newblock \emph{\bibinfo{journal}{\apjl}} \textbf{\bibinfo{volume}{902}}, \bibinfo{pages}{L5} (\bibinfo{year}{2020}).

\bibitem{Huang:2020b}
\bibinfo{author}{{Huang}, C.~X.} \emph{et~al.}
\newblock \bibinfo{title}{{Photometry of 10 Million Stars from the First Two Years of TESS Full Frame Images: Part II}}.
\newblock \emph{\bibinfo{journal}{\rnaas}} \textbf{\bibinfo{volume}{4}}, \bibinfo{pages}{206} (\bibinfo{year}{2020}).

\bibitem{Kovacs:2002}
\bibinfo{author}{{Kov{\'a}cs}, G.}, \bibinfo{author}{{Zucker}, S.} \& \bibinfo{author}{{Mazeh}, T.}
\newblock \bibinfo{title}{{A box-fitting algorithm in the search for periodic transits}}.
\newblock \emph{\bibinfo{journal}{\aap}} \textbf{\bibinfo{volume}{391}}, \bibinfo{pages}{369--377} (\bibinfo{year}{2002}).

\bibitem{Hartman:2012}
\bibinfo{author}{{Hartman}, J.}
\newblock \bibinfo{title}{{VARTOOLS: Light Curve Analysis Program}}.
\newblock \bibinfo{howpublished}{Astrophysics Source Code Library, record ascl:1208.016} (\bibinfo{year}{2012}).

\bibitem{Holczer:2016}
\bibinfo{author}{{Holczer}, T.} \emph{et~al.}
\newblock \bibinfo{title}{{Transit Timing Observations from Kepler. IX. Catalog of the Full Long-cadence Data Set}}.
\newblock \emph{\bibinfo{journal}{\apjs}} \textbf{\bibinfo{volume}{225}}, \bibinfo{pages}{9} (\bibinfo{year}{2016}).

\end{thebibliography}

\begin{thebibliography}{10}
\expandafter\ifx\csname url\endcsname\relax
  \def\url#1{\burl{#1}}\fi
\expandafter\ifx\csname urlprefix\endcsname\relax\def\urlprefix{URL }\fi
\providecommand{\bibinfo}[2]{#2}
\providecommand{\eprint}[2][]{\url{#2}}
\providecommand{\doi}[1]{\url{https://doi.org/#1}}
\bibcommenthead

\bibitem{PS1}
\bibinfo{author}{{Tonry}, J.~L.} \emph{et~al.}
\newblock \bibinfo{title}{{The Pan-STARRS1 Photometric System}}.
\newblock \emph{\bibinfo{journal}{\apj}} \textbf{\bibinfo{volume}{750}}, \bibinfo{pages}{99} (\bibinfo{year}{2012}).

\bibitem{GaiaDR3}
\bibinfo{author}{{Gaia Collaboration}} \emph{et~al.}
\newblock \bibinfo{title}{{Gaia Data Release 3: Summary of the content and survey properties}}.
\newblock \emph{\bibinfo{journal}{arXiv e-prints}} \bibinfo{pages}{arXiv:2208.00211} (\bibinfo{year}{2022}).

\bibitem{Skrutskie2006}
\bibinfo{author}{{Skrutskie}, M.~F.} \emph{et~al.}
\newblock \bibinfo{title}{{The Two Micron All Sky Survey (2MASS)}}.
\newblock \emph{\bibinfo{journal}{\aj}} \textbf{\bibinfo{volume}{131}}, \bibinfo{pages}{1163--1183} (\bibinfo{year}{2006}).

\bibitem{Wright2010}
\bibinfo{author}{{Wright}, E.~L.} \emph{et~al.}
\newblock \bibinfo{title}{{The Wide-field Infrared Survey Explorer (WISE): Mission Description and Initial On-orbit Performance}}.
\newblock \emph{\bibinfo{journal}{\aj}} \textbf{\bibinfo{volume}{140}}, \bibinfo{pages}{1868--1881} (\bibinfo{year}{2010}).

\bibitem{ariadne}
\bibinfo{author}{{Vines}, J.~I.} \& \bibinfo{author}{{Jenkins}, J.~S.}
\newblock \bibinfo{title}{{ARIADNE: measuring accurate and precise stellar parameters through SED fitting}}.
\newblock \emph{\bibinfo{journal}{\mnras}} \textbf{\bibinfo{volume}{513}}, \bibinfo{pages}{2719--2731} (\bibinfo{year}{2022}).

\bibitem{Husser2013}
\bibinfo{author}{Husser, T.-O.} \emph{et~al.}
\newblock \bibinfo{title}{{Astrophysics A new extensive library of PHOENIX stellar atmospheres}}.
\newblock \emph{\bibinfo{journal}{A{\&}A}} \textbf{\bibinfo{volume}{553}}, \bibinfo{pages}{A6} (\bibinfo{year}{2013}).

\bibitem{Allard2011}
\bibinfo{author}{{Allard}, F.}, \bibinfo{author}{{Homeier}, D.} \& \bibinfo{author}{{Freytag}, B.}
\newblock \bibinfo{editor}{{Johns-Krull}, C.}, \bibinfo{editor}{{Browning}, M.~K.} \& \bibinfo{editor}{{West}, A.~A.} (eds) \emph{\bibinfo{title}{{Model Atmospheres From Very Low Mass Stars to Brown Dwarfs}}}.
\newblock (eds \bibinfo{editor}{{Johns-Krull}, C.}, \bibinfo{editor}{{Browning}, M.~K.} \& \bibinfo{editor}{{West}, A.~A.}) \emph{\bibinfo{booktitle}{16th Cambridge Workshop on Cool Stars, Stellar Systems, and the Sun}}, Vol. \bibinfo{volume}{448} of \emph{\bibinfo{series}{Astronomical Society of the Pacific Conference Series}}, \bibinfo{pages}{91} (\bibinfo{year}{2011}).

\bibitem{KuruczModel}
\bibinfo{author}{{Kurucz}, R.}
\newblock \bibinfo{title}{{ATLAS9 Stellar Atmosphere Programs and 2 km/s grid.}}
\newblock \emph{\bibinfo{journal}{ATLAS9 Stellar Atmosphere Programs and 2 km/s grid. Kurucz CD-ROM No. 13. Cambridge}} \textbf{\bibinfo{volume}{13}} (\bibinfo{year}{1993}).

\bibitem{Castelli2004}
\bibinfo{author}{{Castelli}, F.} \& \bibinfo{author}{{Kurucz}, R.~L.}
\newblock \bibinfo{editor}{{Piskunov}, N.}, \bibinfo{editor}{{Weiss}, W.~W.} \& \bibinfo{editor}{{Gray}, D.~F.} (eds) \emph{\bibinfo{title}{{New Grids of ATLAS9 Model Atmospheres}}}.
\newblock (eds \bibinfo{editor}{{Piskunov}, N.}, \bibinfo{editor}{{Weiss}, W.~W.} \& \bibinfo{editor}{{Gray}, D.~F.}) \emph{\bibinfo{booktitle}{Modelling of Stellar Atmospheres}}, Vol. \bibinfo{volume}{210}, \bibinfo{pages}{A20} (\bibinfo{year}{2003}).
\newblock \eprint{astro-ph/0405087}.

\bibitem{Gaia}
\bibinfo{author}{{Gaia Collaboration}} \emph{et~al.}
\newblock \bibinfo{title}{{The Gaia mission}}.
\newblock \emph{\bibinfo{journal}{\aap}} \textbf{\bibinfo{volume}{595}}, \bibinfo{pages}{A1} (\bibinfo{year}{2016}).

\bibitem{DR3ParallaxCorrection}
\bibinfo{author}{{Lindegren}, L.} \emph{et~al.}
\newblock \bibinfo{title}{{Gaia Early Data Release 3. Parallax bias versus magnitude, colour, and position}}.
\newblock \emph{\bibinfo{journal}{\aap}} \textbf{\bibinfo{volume}{649}}, \bibinfo{pages}{A4} (\bibinfo{year}{2021}).

\bibitem{radvel}
\bibinfo{author}{{Fulton}, B.~J.}, \bibinfo{author}{{Petigura}, E.~A.}, \bibinfo{author}{{Blunt}, S.} \& \bibinfo{author}{{Sinukoff}, E.}
\newblock \bibinfo{title}{{RadVel: The Radial Velocity Modeling Toolkit}}.
\newblock \emph{\bibinfo{journal}{\pasp}} \textbf{\bibinfo{volume}{130}}, \bibinfo{pages}{044504} (\bibinfo{year}{2018}).

\end{thebibliography}
\end{document}